\documentclass[10pt]{article}

\usepackage[utf8]{inputenc}
\usepackage[]{natbib}

\usepackage{authblk}
\usepackage{setspace}

\usepackage{subfigure}

\usepackage{amsmath}
\usepackage{amssymb}
\usepackage{latexsym}
\usepackage{bm}
\usepackage{multirow}

\usepackage{ifpdf}

\ifpdf
\usepackage[pdftex]{graphicx}
\else
\usepackage{graphicx}
\fi

\usepackage{algorithm, algorithmic}
\usepackage{color}

\title{An Adaptive Interacting Wang-Landau Algorithm\\ for Automatic Density Exploration}

\author[1]{Luke Bornn}
\author[2]{Pierre E. Jacob}
\author[3]{Pierre Del Moral}
\author[4]{Arnaud Doucet}
\affil[1]{Department of Statistics, Harvard University, USA}
\affil[2]{CEREMADE, Universit\'{e} Paris-Dauphine, France}
\affil[3]{INRIA Bordeaux Sud-Ouest and University of Bordeaux, France}
\affil[4]{Department of Statistics, Oxford University, UK}
\RequirePackage[colorlinks,citecolor=blue,urlcolor=blue]{hyperref}

\begin{document}

\ifpdf
\DeclareGraphicsExtensions{.pdf, .png, .jpg, .tif}
\else
\DeclareGraphicsExtensions{.eps, .jpg}
\fi

\maketitle

\begin{abstract}
While statisticians are well-accustomed to performing exploratory analysis in
the modeling stage of an analysis, the notion of conducting preliminary
general-purpose exploratory analysis in the Monte Carlo stage (or more
generally, the model-fitting stage) of an analysis is an area which we feel
deserves much further attention.  Towards this aim, this paper proposes a
general-purpose algorithm for automatic density exploration.  The proposed
exploration algorithm combines and expands upon components from various
adaptive Markov chain Monte Carlo methods, with the Wang-Landau algorithm at
its heart.  Additionally, the algorithm is run on interacting parallel chains
-- a feature which both decreases computational cost as well as stabilizes the
algorithm, improving its ability to explore the density.  Performance of this
new parallel adaptive Wang-Landau (PAWL) algorithm is studied in several
applications.  Through a Bayesian variable selection example, the authors
demonstrate the convergence gains obtained with interacting chains.  The
ability of the algorithm's adaptive proposal to induce mode-jumping is
illustrated through a Bayesian mixture modeling application.  Lastly, through a
2D Ising model, the authors demonstrate the ability of the algorithm to
overcome the high correlations encountered in spatial models. The appendices
contain the full algorithmic description in pseudo-code, a tri-modal toy example and 
remarks on the convergence of the proposed algorithm.
\end{abstract}

\section{Introduction}

As improvements in technology introduce measuring devices capable of capturing
ever more complex real-world phenomena, the accompanying models used to
understand such phenomena grow accordingly.  While linear models under the
assumption of Gaussian noise were the hallmark of early 20th century
statistics, the past several decades have seen an explosion in statistical
models which produce complex and high-dimensional density functions for which
simple, analytical integration is impossible.  This growth was largely fueled
by renewed interest in Bayesian statistics accompanying the Markov chain Monte
Carlo (MCMC) revolution in the 1990's.  With the computational power to explore
the posterior distributions arising from Bayesian models, MCMC allowed
practitioners to build models of increasing size and nonlinearity.

As a core component of many of the MCMC algorithms discussed later, we briefly
recall the Metropolis-Hastings algorithm. With the goal of sampling from a
density $\pi$, the algorithm generates a Markov chain $(x_t)_{t =1}^T$ with
invariant distribution $\pi$. From a current state $x_t$, a new state $x'$ is
sampled using a proposal density $q_{\eta}(x_t, x')$ parametrized by $\eta$.
The proposed state $x'$ is accepted as the new state $x_{t+1}$ of the chain
with probability
\begin{align*}
	\min \left( 1 , \frac{\pi(x')q_{\eta}( x', x_t)}{\pi(x_t)q_{\eta}( x_t, x')} \right)
\end{align*}
and if it is rejected, the new state $x_{t+1}$ is set to the previous state
$x_t$. From this simple algorithmic description, it is straightforward to see
that if $x_t$ is in a local mode and the proposal density $q_{\eta}(x_t, x')$
has not been carefully chosen to propose samples from distant regions, the
chain will become stuck in the current mode. This is due to the rejection of
samples proposed outside the mode, underscoring the importance of ensuring
$q_{\eta}(x_t, x')$ is intelligently designed.

Though standard MCMC algorithms such as the Metropolis-Hastings algorithm and
the Gibbs sampler have been studied thoroughly and the convergence to the
target distribution is ensured under weak assumptions, many applications
introduce distributions which cannot be sampled easily by these algorithms.
Multiple reasons can lead to failure in practice, even if long-run convergence
is guaranteed; the question then becomes whether or not the required number of
iterations to accurately approximate the density is reasonable given the
currently available computational power. Among these reasons, let us cite a few
that will be illustrated in later examples: the probability density function
might be highly multimodal, in which case the chain can get stuck in local
modes. Alternatively or additionally, it might be defined on a high-dimensional
state space with strong correlations between the components, in which case the
proposal distributions (and in particular their large covariance matrices) are
very difficult to tune manually. These issues lead to error and bias in the
resulting inference, and may be detected through convergence monitoring
techniques (see, e.g., \citet{Robert2004a}).  However, even when convergence is
monitored, it is possible that entire modes of the posterior are missed.  To
address these issues, we turn to a burgeoning class of Monte Carlo methods
which we refer to as ``exploratory algorithms.''

In the following section, we discuss the traits that allow exploratory MCMC
algorithms to perform inference in multimodal, high-dimensional distributions,
connecting these traits to existing exploratory algorithms in the process.  In
Section 3, we detail one of these, the Wang-Landau algorithm, and propose
several novel improvements that make it more adaptive, hence easier to use, and
also improve convergence.  Section 4 applies the proposed algorithm to variable
selection, mixture modeling and spatial imaging, before Section 5 concludes.

\section{Exploratory Algorithms \label{sec:algocomponents}}

As emphasized by \citet{Schmidler2011a}, there are two distinct goals of
existing adaptive algorithms.  Firstly, algorithms which adapt the proposal
according to past samples are largely exploitative, in that they improve
sampling of features already seen.  However, modes or features not yet seen by
the sampler might be quite different from the previously explored region, and
as such adaptation might prevent adequate exploration of alternate regions of
the state space.  As an attempted solution to this problem
\citet{LearnFromThyNeighbor} suggest adapting regionally, with parallel chains
used to perform the adaptation.  Secondly, there exists a set of adaptive
algorithms whose goal is to adapt in such a way as to encourage density
exploration.  These include, for instance, the equi-energy sampler
\citep{Kou2005a}, parallel tempering \citep{Swendsen1986a, Geyer1991a}, and the
Wang-Landau \citep{Wang2001a, Wang2001b, Liang2005a, Atchade2010a} algorithms
among others.  The algorithm developed here fits into the latter suite of
tools, whose goal is to explore the target density, particularly distant and
potentially unknown modes.

Although the aforementioned algorithms have proven efficient for specific
challenging inference problems, they are not necessarily designed to be
generic, and it is often difficult and time-consuming for practitioners to
learn and code these algorithms merely to test a model.  As such, while
statisticians are accustomed to exploratory data analysis, we believe that
there is room for generic exploratory Monte Carlo algorithms to learn the basic
features of the distribution or model of interest, particularly the locations
of modes and regions of high correlation. These generic algorithms would
ideally be able to deal with discrete and continuous state spaces and any
associated distribution of interest, and would require as few parameters to
tune as possible, such that users can use them before embarking on
time-consuming, tailor-made solutions designed to estimate expectations with
high precision.  In this way one may perform inference and compare between a
wide range of models without building custom-purpose Monte Carlo methods for
each.

We first describe various ideas that have been used to explore
highly-multimodal densities, and then describe recent works aimed at
automatically tuning algorithmic parameters of MCMC methods, making them able
to handle various situations without requiring much case-specific work from the
user.

\subsection{Ability to Cross Low-Density Barriers \label{sec:crossbarriers}}

The fundamental problem of density exploration is settling into local modes,
with an inability to cross low-density regions to find alternative modes.  For
densities $\pi$ which are highly multi-modal, or ``rugged,'' one can employ
tempering strategies, sampling instead from a distribution proportional to
$\pi^{1/\tau}$ with temperature $\tau>1$.  Through tempering, the peaks and
valleys of $\pi$ are smoothed, allowing easier exploration.  This is the
fundamental idea behind parallel tempering, which employs multiple chains at
different temperatures; samples are then swapped between chains, using highly
tempered chains to assist in the exploration of the untempered chain
\citep{Geyer1991a}.  \citet{Marinari1992a} subsequently proposed simulated
tempering, which dynamically moves a single chain up or down the temperature
ladder. One may also fit tempering within a sequential Monte Carlo approach,
whereby samples are first obtained from a highly tempered distribution; these
samples are transitioned through a sequence of distributions converging to
$\pi$ using importance sampling and moves with a Markov kernel
\citep{Neal2001a, Del-Moral2006a}. However, using tempering strategies with
complex densities, one must be careful of phase transitions, where the density
transforms considerably across a given temperature value. 

A related class of algorithms works by partitioning the state space along the
energy function $-\log \pi(x)$.  The idea of slicing, or partitioning, along
the energy function is the hallmark of several auxiliary variable sampling
methods, which iteratively sample $U \sim \mathcal{U}[0, \pi(X)]$ then $X \sim
\mathcal{U}\{X:\pi(X) \geq U\}$.  This is the fundamental idea behind the
Swendsen–Wang algorithm \citep{Swendsen1987a, Edwards1988a} and related
algorithms (e.g. \citet{Besag1993a, Higdon1998b, Neal2003a}).  The equi-energy
sampler \citep{Kou2005a, Baragatti2012a}, in contrast to the above auxiliary
variable methods, begins by sampling from a highly tempered distribution; once
convergence is reached, a new reduced-temperature chain is run with updates
from a mixture of Metropolis moves and exchanges of the current state with the
value of a previous chain in the same energy band.  The process is continued
until the temperature reaches $1$ and the invariant distribution of the chain
is the target of interest.  As such, this algorithm works through a sequence of
tempered distributions, using previous distributions to create intelligent
mode-jumping moves along an equal-energy set.

In a similar vein, the Wang-Landau algorithm \citep{Wang2001a,Wang2001b} also
partitions the state space $\mathcal{X}$ along a reaction coordinate $\xi(x)$,
typically the energy function: $\xi(x) = -\log \pi(x)$, resulting in a
partition  $(\mathcal{X}_i)_{i=1}^d$. The algorithm generates a
time-inhomogeneous Markov chain that admits an invariant distribution
$\tilde{\pi}_t$ at iteration $t$, instead of the target distribution $\pi$
itself as e.g. in a standard Metropolis-Hastings algorithm. The distribution
$\tilde{\pi}_t$ is designed such that the generated chain equally visits the
various regions $\mathcal{X}_i$ as $t \rightarrow \infty$. Because the
Wang-Landau algorithm lies at the heart of our proposed algorithm, it is
extensively described in Section \ref{sec:proposedalgorithm}.

It is worth discussing a similar, recently proposed algorithm which combines
Markov chain Monte Carlo and free energy biasing \citep{ChopinLelievreStoltz}
and its sequential Monte Carlo counterpart \citep{Chopin2010b}. The central
idea of the latter is to explore a sequence of distributions, successively
biasing according to a reaction coordinate $\xi$ in a similar manner.  However,
we have found the method to be largely dependent on selecting a well-chosen
initial distribution $\pi_0$, as is usually the case with sequential Monte
Carlo methods for static inference. If the initial distribution is not chosen
to be flatter than the target distribution, which is possibly the case since
the regions of interest with respect to the target distribution are \emph{a
priori} unknown, the efficiency of the SMC methods relies mostly on the move
steps within the particle filter, which are themselves Metropolis--Hastings or
Gibbs moves.

\subsection{Adaptive Proposal Mechanism}

Concurrent with the increasing popularity of exploratory methods, the issue of
adaptively fine-tuning MCMC algorithms has also seen considerable growth since
the foundational paper of \citet{Haario2001a}, including a series of
conferences dedicated solely to the problem (namely, Adap'ski 1 through 3 among
others); see the reviews of \citet{Andrieu2008b} and  \citet{Atchade2009c} for
more details.  While the de-facto standard has historically been hand-tuning of
MCMC algorithms, this new work finds interest in automated tuning, resulting in
a new class of methods called adaptive MCMC.

The majority of the existing literature focuses on creating intelligent
proposal distributions for an MCMC sampler.  The principal idea is to exploit
past samples to induce better moves across the state space by matching moments
of the proposal and past samples, or by encouraging a particular acceptance
rate of the sampler.  The raison d'\^{e}tre of these algorithms is that tuning
MCMC algorithms by hand is both time-consuming and prone to inaccuracies.  By
automating the selection of the algorithm's parameters, practitioners might
save considerable time in their analyses.  This feature is pivotal in an
automated density exploration algorithm.  Due to its exploratory nature, it is
likely that the practitioner might not have complete knowledge of even the
scale of the density support; as a result, having a proposal distribution which
adapts to the density at hand is a crucial step in the automation process.

One must be careful in selecting the type of adaptation mechanism employed to
encourage exploration, rather than simply exploiting previously explored modes.
For instance, tuning a proposal covariance to a previously visited mode might
prevent the algorithm from reaching as yet unexplored modes in the direction of
the current mode's minor axis.  Additionally, when combined with a
progressively biased distribution as in the Wang-Landau algorithm, it is
desirable to have a proposal which first samples what it sees well, then later
grows in step size to better explore the flattened (biased) distribution.

\section{Proposed Algorithm \label{sec:proposedalgorithm}}

We now develop our proposed algorithm. After recalling the Wang-Landau
algorithm, which constitutes the core of our method, we describe three
improvements: an adaptive binning strategy to automate the difficult task of
partitioning the state space, the use of interacting parallel chains to improve
the convergence speed and use of computational resources, and finally the use
of adaptive proposal distributions to encourage exploration as well as to
reduce the number of algorithmic parameters. We detail at the end of the
section how to use the output of the algorithm, which we term parallel adaptive
Wang-Landau (PAWL) to answer the statistical problem at hand.

\subsection{The Wang-Landau Algorithm}

\begin{figure}[H]
\centering
\includegraphics[width=0.95\textwidth]{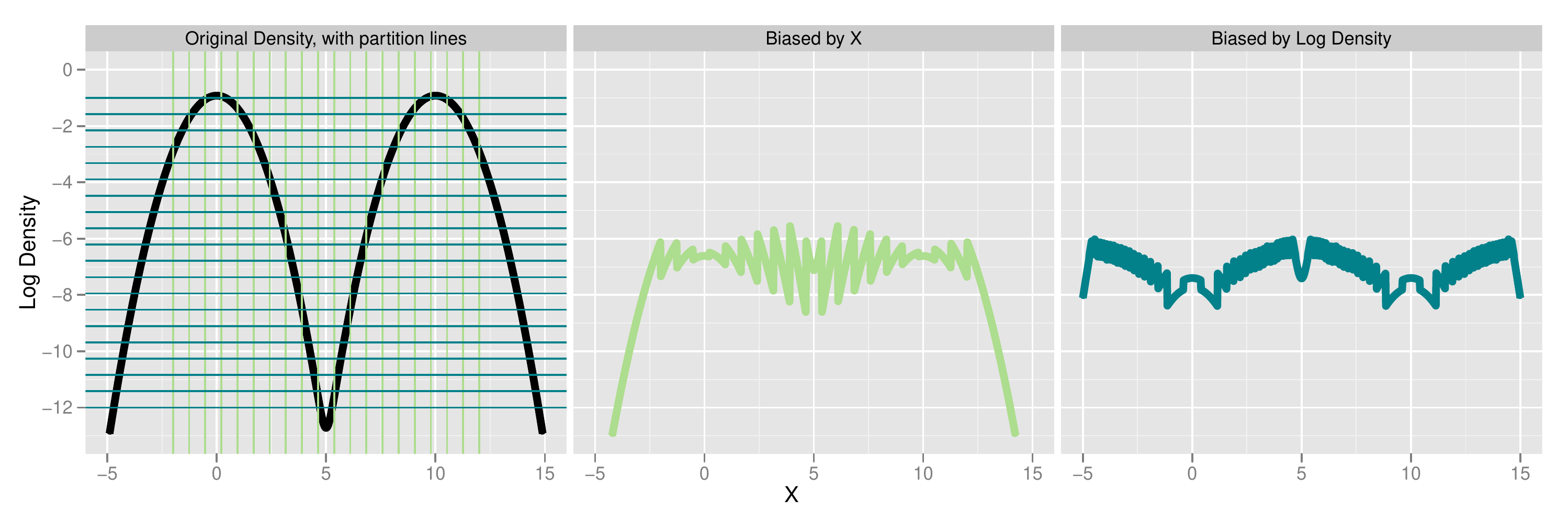}
\caption{\label{fig:biastoy} Probability density functions for a univariate distribution $\pi$ and its biased version $\tilde{\pi}$ when partitioning the state space along the $x$-axis ($\xi(x) = x$, middle) and the log density ($\xi(x) = -\log \pi(x)$, right).  The left-most plot also shows the partitioning of the state space with $\xi(x)$; in both cases $d = 20$. The biasing is done such that the integral $\int_{\mathcal{X}_i} \tilde{\pi}(x) \mathrm{d}x$ is the same for all $\mathcal{X}_i$ (areas under the curve for each set) and such that $\pi$ and $\tilde{\pi}$ coincide on each set $\mathcal{X}_i$, up to a multiplicative constant.}
\end{figure}

As previously mentioned, the Wang-Landau algorithm generates a
time-inhomogeneous Markov chain that admits a distribution $\tilde{\pi}_t$ as
the invariant distribution at iteration $t$.  The biased distribution
$\tilde{\pi}_t$ targeted by the algorithm at iteration $t$ is based on the
target distribution $\pi$, and modified such that a) the generated chain visits
all the sets $(\mathcal{X}_i)_{i=1}^d$ equally, that is the proportion of
visits in each set is converging to $d^{-1}$ when $t$ goes to infinity; and b)
the restriction of the modified distribution $\tilde{\pi}_t$ to each set
$\mathcal{X}_i$ coincides with the restriction of the target distribution $\pi$
to this set, up to a multiplicative constant.  The modification (a) is crucial,
as inducing uniform exploration of the sets is the biasing mechanism which
improves exploration; in fact similar strategies are used in other fields,
including combinatorial optimization \citep{wei2004towards}. Ideally the biased
distribution $\tilde{\pi}$ would not depend on $t$, and would be available
analytically as:
\begin{align}
  \tilde{\pi}(x) = \pi(x) \times \frac{1}{d} \sum_{i=1}^{d} \frac{\mathcal{I}_{\mathcal{X}_i}(x)}{\psi(i)} 
	\label{BiasedDensity}
\end{align}
where $\psi(i) = \int_{\mathcal{X}_i} \pi(x) \mathrm{d}x$ and
$\mathcal{I}_{\mathcal{X}_i}(x)$ is equal to $1$ if $x\in\mathcal{X}_i$ and 0
otherwise. Checking that using $\tilde{\pi}$ as the invariant distribution of a
MCMC algorithm would validate points a) and b) is straightforward. Figure
\ref{fig:biastoy} illustrates a univariate target distribution $\pi$ and its
corresponding biased distribution $\tilde{\pi}$ under two different partitions
of the state space.

In practical situations, however, the integrals $(\psi(i))_{i=1}^d$ are not
available, hence we wish to plug estimates $(\theta(i))_{i=1}^d$ of
$(\psi(i))_{i=1}^d$ into Equation \eqref{BiasedDensity}. The Wang-Landau
algorithm is an iterative algorithm which jointly generates a sequence of
estimates $(\theta_t(i))_t$ for all $i$ and a Markov chain $(X_t)_t$, such that
when $t$ goes to infinity, $\theta_t(i)$ converges to $\psi(i)$ and
consequently, the distribution of $X_t$ converges to $\tilde{\pi}$. We denote
by $\tilde{\pi}_{\theta_t}$ the biased distribution obtained by replacing
$\psi(i)$ by its estimate $\theta_t(i)$ in Equation \eqref{BiasedDensity}. Note
that the normalizing constant of $\tilde{\pi}_{\theta_t}$ is now unknown. A
simplified version of the Wang-Landau algorithm is given in Algorithm
\ref{algo:WL}.

\begin{algorithm}
\caption{Simplified Wang-Landau Algorithm\label{algo:WL}}
\begin{algorithmic}[1]
\STATE Partition the state space into $d$ regions $\{ \mathcal{X}_1, \dots,
\mathcal{X}_d \}$ along a reaction coordinate $\xi(x)$. 
\STATE First, $\forall i\in \{ 1,\dots, d \}$ set $\theta(i) \leftarrow 1$.
\STATE Choose a decreasing sequence $\{ \gamma_t \}$, typically $\gamma_t = 1/t$.
\STATE Sample $X_0$ from an initial distribution $\pi_0$.
\FOR {$t=1$ to $T$}
      \STATE Sample $X_t$ from $P_{\theta_{t-1}}(X_{t-1},\cdot)$, a
transition kernel with invariant distribution $\tilde{\pi}_{\theta_{t-1}}(x)$.
      \STATE Update the bias: $\log \theta_{t}(i) \leftarrow \log  \theta_{t-1}(i) + \gamma_t
(\mathcal{I}_{\mathcal{X}_i}(X_t) - d^{-1})$.
      \STATE Normalize the bias: $\theta_t(i) \leftarrow \theta_t(i) / \sum_{i = 1}^d \theta_t(i)$.
\ENDFOR
\end{algorithmic}
\end{algorithm}

The rationale behind the update of the bias is that if the chain is in the set
$\mathcal{X}_i$, the probability of remaining in $\mathcal{X}_i$ should be
reduced compared to the other sets through an increase in the associated bias
$\theta_t(i)$. Therefore the chain is pushed towards the sets that have been
visited less during the previous iterations, improving the exploration of the
state space so long as the partition $(\mathcal{X}_i)_{i=1}^d$ is well chosen.
While this biasing mechanism adds cost to each iteration of the algorithm the
tradeoff is improved exploration.  In step $6$, the transition kernel is
typically a Metropolis-Hastings move, due to the lack of conjugacy brought
about by biasing.

In this simplified form the Wang-Landau algorithm reduces to standard
stochastic approximation, where the term $\gamma_t$ decreases at each
iteration. The algorithm as given in \citet{Wang2001a,Wang2001b} uses a more
sophisticated learning rate $\gamma_t$ which does not decrease
deterministically, but instead only when a certain criterion is met. This
criterion, referred to as the ``flat histogram'' criterion, is met when for all
$i \in \{ 1,\dots, d \}$, $\nu(i)$ is close enough to $d^{-1}$, where we denote
by $\nu(i)$ the proportion of visits of $(X_t)$ in the set $\mathcal{X}_i$
since the last time the criterion was met.  Hence we introduce a real number
$c$ to control the distance between $\nu(i)$ and $d^{-1}$, and an integer $k$
to count the number of criteria already met.  We describe the generalized
Wang-Landau in Algorithm \ref{algo:generalizedWL}.

\begin{algorithm}
\caption{Wang-Landau Algorithm\label{algo:generalizedWL}}
\begin{algorithmic}[1]
\STATE Partition the state space into $d$ regions $\{ \mathcal{X}_1, \dots,
\mathcal{X}_d \}$ along a reaction coordinate $\xi(x)$.
\STATE First, $\forall i\in \{ 1, \dots, d \}$ set $\theta(i)\leftarrow1,  \nu(i) \leftarrow 0$.
\STATE Choose a decreasing sequence $\{ \gamma_k \}$, typically $\gamma_k = 1/k$.
\STATE Sample $X_0$ from an initial distribution $\pi_0$. \label{algo:generalizedWL:init}
\FOR {$t=1$ to $T$}
      \STATE Sample $X_t$ from $P_{\theta_{t-1}}(X_{t-1},\cdot)$, a
transition kernel with invariant distribution $\tilde{\pi}_{\theta_{t-1}}(x)$. \label{algo:generalizedWL:propagate}
      \STATE Update the proportions: $\forall i\in \{1, \dots, d\} \quad \nu(i) \leftarrow \frac{1}{t}\left[ (t-1) \nu(i) + \mathcal{I}_{\mathcal{X}_i}(X_t)\right]$. \label{algo:generalizedWL:nu}
      \IF {``flat histogram'': $\max_{i \in [1,d]} \vert \nu(i) - d^{-1}\vert < c/d$}
	  \STATE Set $k \leftarrow k + 1$.
	  \STATE Reset $\forall i\in \{1,\dots, d\} \quad \nu(i) \leftarrow 0$.
      \ENDIF 
      \STATE Update the bias: $\log \theta_{t}(i) \leftarrow \log  \theta_{t-1}(i) + \gamma_k
(\mathcal{I}_{\mathcal{X}_i}(X_t) - d^{-1})$.\label{algo:generalizedWL:bias}
      \STATE Normalize the bias: $\theta_t(i) \leftarrow \theta_t(i) / \sum_{i = 1}^d \theta_t(i)$.
\ENDFOR
\end{algorithmic}
\end{algorithm}

When $c$ is set to low values (e.g. $c = 0.1$ or $0.5$), the algorithm must
explore the various regions such that the frequency of visits to the region
$\mathcal{X}_i$ is approximately $d^{-1}$ before the learning rate $\gamma_k$
is decreased.  Also, the algorithm may be further generalized to target a
desired frequency $\phi_i$ instead of the same frequency $d^{-1}$ for every
set; while such strategies may be useful, as demonstrated in the following
section, for notational simplicity we focus on the case $\phi_i = d^{-1}$.  As
already mentioned, to answer the general question of exploring the support of a
target density $\pi$, the default choice for the reaction coordinate is the
energy function: $\xi(x) = -\log \pi(x)$, which has the benefit of being
one-dimensional regardless of the dimension of the state space $\mathcal{X}$.
However, for specific models other reaction coordinates have been used, such as
one (or more) of the components $x_j$ of $x$ or a linear combination of
components of $x$.  In the applications in Section \ref{section:ref} we discuss
the use of alternative reaction coordinates further.

We now propose improvements to the Wang-Landau algorithm to increase its
flexibility and efficiency.

\subsection{A Novel Adaptive Binning Strategy}

The Wang-Landau and equi-energy sampler algorithms are known to perform well if
the bins, or partitions of the one-dimensional reaction coordinate $\xi(x)$,
are well chosen. However, depending on the problem it might be difficult to
choose the bins to optimize sampler performance. A typical empirical approach
to deal with this issue is to first run, for example, an adaptive MCMC
algorithm to find at least one mode of the target distribution. The generated
sample and the associated target density evaluations determine a first range of
the target density values which can be used to initialize the bins. At this
point the user can choose a wider range of target density values (e.g. by
multiplying the range by 2), in order to allow for a wider exploration of the
space. Within this initial range, one must still decide the number of bins.

\begin{figure}
    \centering
    \subfigure[Unique bin before the split]{\label{subfig:beforesplit}\includegraphics[width=0.46\textwidth]{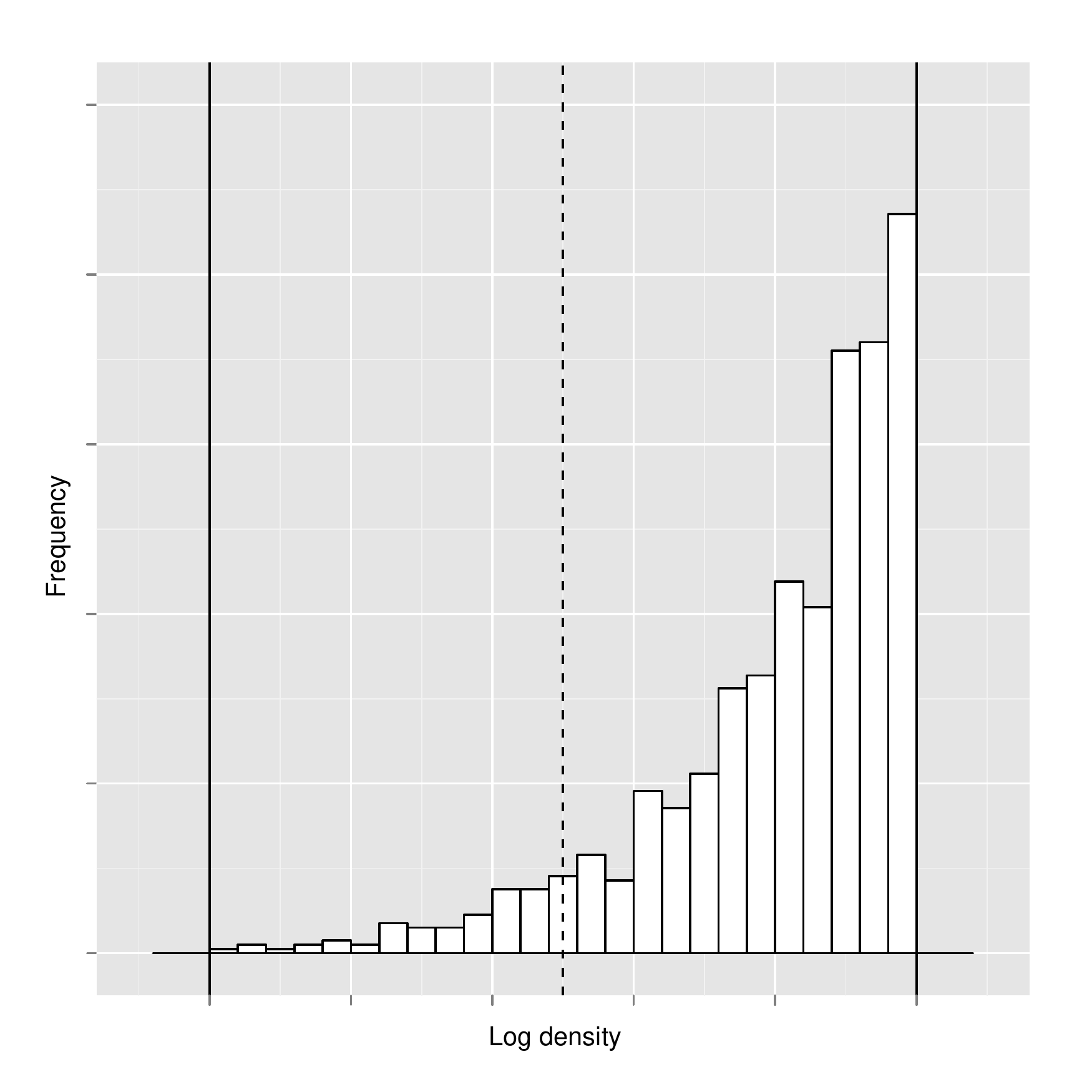}}
    \subfigure[Two bins after the split]{\label{subfig:aftersplit} \includegraphics[width=0.46\textwidth]{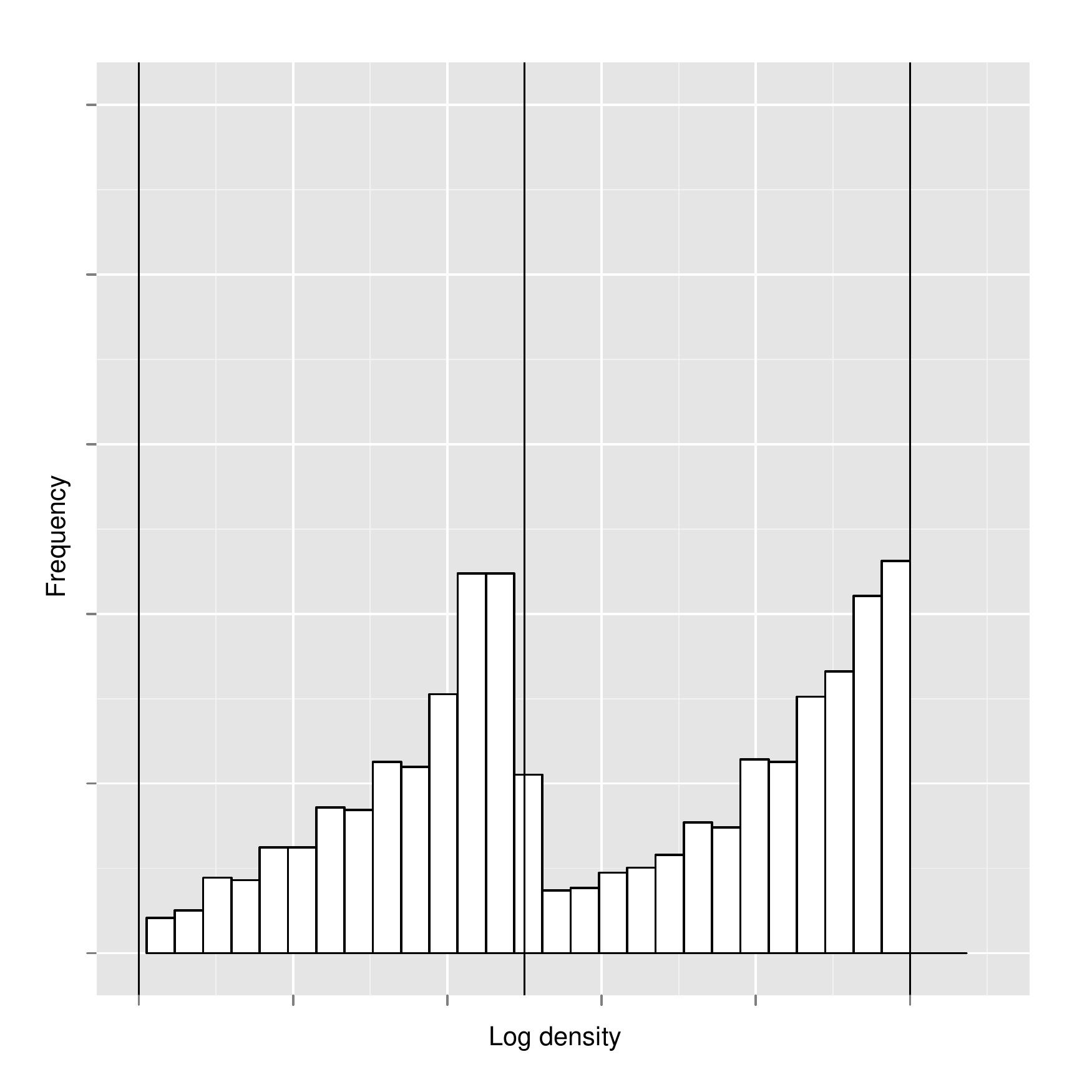}}
    \caption{\label{fig:automaticsplit} Artificial histograms of the log target
    density values associated to the chains generated by the algorithm, within a
    single bin (left) and within two bins created by splitting the former bin
    (right).}
\end{figure}

Due to difficulties with selecting the bins, it has been suggested that one
should adaptively compare adjacent bins, splitting a bin if the corresponding
estimate $\theta$ is significantly larger than a neighboring value
\citep{Schmidler2011a}.  
Because each $\psi_i$ is a given bin's normalizing constant, we feel it is more
important to maintain uniformity within a bin to allow easy within-bin
movement.  Our proposed approach to achieve this ``flatness'' is to look at the
distribution of the realized reaction coordinate values within each bin. Figure
\ref{fig:automaticsplit} illustrates this distribution on an artificial
histogram. The plot of Figure \ref{subfig:beforesplit} shows a situation where,
within one bin, the distribution might be strongly skewed towards one side. In
this artificial example, very few points have visited the left side of the bin,
which suggests that moving from this bin to the left neighboring bin might be
difficult. 

We propose to consider the ratio of the number of points on the left side of
the middle (dashed line) over the number of points within the bin as a very
simple proxy for the discrepancy of the chains within one bin (see e.g.
\citet{Niederreiter1992a} for much more sophisticated discrepancy measures). In
a broad outline, if this ratio was around $50\%$, the within-bin histogram
would be roughly uniform. On the contrary, the ratio corresponding to Figure
\ref{subfig:beforesplit} is around $7\%$.  Our strategy is to split the bin if
this ratio goes below a given threshold, say $25\%$; two new bins are created,
corresponding to the left side and the right side of the former bin, and each
bin is assigned a weight of $\theta/2$ where $\theta$ is the weight of the
former bin. These provide starting values for the estimation of the weight of
the new bins during the following iterations of the algorithm.  Note also that
the desired frequency of visits to each of the new bins, which was for instance
equal to $1/d$ before the split, has to be specified as well. In the numerical
experiments, we set the desired frequency of the new bins as one half of the
desired frequency of the former bin.  Figure \ref{subfig:aftersplit} shows the
distribution of samples within the two new bins. The resulting histogram is not
uniform, yet exhibits a more even distribution within the bin -- a feature
which is expected to help the chain to move from this bin to the left
neighboring bin.  The threshold could be set closer to $50\%$, which would
result in more splits and therefore more bins.

In practice it is not necessary to check whether the bins have to be split at
every iteration. Our strategy is to check every $n$-th iteration, until the
flat histogram criterion is met for the first time. When it is met, it means
that the chains can move easily between the bins, and hence the bins can be
kept unchanged for the remaining iterations. Finally, when implementing the
automatic binning strategy for discrete distributions, one must ensure that a
new bin corresponds to actual points in the state space. For example if the
bins are along the energy values and the state space is finite, there are
certainly intervals of energy to which no states corresponds, and that would
therefore never be reached.  Section 4 demonstrates the proposed adaptive
binning strategy in practice.

In addition to allowing for splitting of bins, it is also important to allow
the range of bins to extend if new states are found outside of the particular
range.  That said, one must
differentiate between the two extremes of the reaction coordinate.  For
example, if $\xi(x) = -\log \pi(x)$, then one might not wish to add more
low-density (high-energy) bins, which would induce the sampler to explore
further and further into the tails.  However, if one finds a new high-density
(low-energy) mode beyond the energy range previously seen, then the sampler
might become stuck in this new mode. In this case, we propose to extend the first
bin corresponding to the lowest level of energy to always include the lowest 
observed values. The adaptive partition $(\mathcal{X}_{i, t})_{i=1}^{d_t}$ of the state space takes the following form
at time $t$:
\[
\mathcal{X}_{1,t} = [e_{\min, t}, e_{1,t}], \; 
\mathcal{X}_{2,t} = [e_{1,t}, e_{2,t}], \; \ldots
\mathcal{X}_{d_t,t}  = [e_{d_t-1,t}, +\infty)
\]
where $e_{\min, t} = \min_{t\geq 0}\{- \log \pi(X_t)\}$ and $(e_{1,t}, \ldots, e_{d_t-1, t})$
defines the limits of the inner bins at time $t$, which is the result of initial
bin limits $(e_{1,0}, \ldots, e_{d_0-1, 0})$,
and possible splits between time $0$ and time $t$.  As such, if new low-energy values are found, the bin $\mathcal{X}_{1,t}$ is widened. If this results in unequal exploration across the reaction coordinate, then the adaptive bin-splitting mechanism will automatically split this newly widened bin.

\subsection{Parallel Interacting Chains \label{sec:parallel}}

We propose to generate multiple chains instead of a single one to improve
computational scalability through parallelization as well as particle
diversity.  The use of interacting chains has become of much interest in recent
years, with the multiple chains used to create diverse proposals
\citep{casarin2011interacting},  to induce long-range equi-energy jumps
\citep{Kou2005a}, and to generally improve sampler performance; see
\citet{Atchade2009c}, \citet{brockwell2010sequentially}, and
\citet{byrd2010parallel} for recent developments.  The use of parallelization
is not constrained to multiple chains, however, and has also been employed to
speed up the generation of a single chain through pre-fetching
\citep{brockwell2006parallel}.

Let $N$ be the desired number of chains. We follow Algorithm
\ref{algo:generalizedWL}, with the following modifications. First we generate
$N$ starting points $\bm{X}_0 = (X_0^{(1)}, \ldots, X_0^{(N)})$ independently
from an initial distribution $\pi_0$ (Algorithm \ref{algo:generalizedWL} line
\ref{algo:generalizedWL:init}).  Then at iteration $t$, instead of moving one
chain using the transition kernel $P_{\theta_{t-1}}$, we move the $N$ chains
using the same transition kernel, associated with the same bias $\theta_{t-1}$
(Algorithm 2 line \ref{algo:generalizedWL:propagate}). We emphasize that the
bias $\theta$ is common to all chains, which makes the proposed method
different from running Wang-Landau chains entirely in parallel.  The
proportions $\nu(i)$ are updated using all the chains, simply by replacing the
indicator function $\mathcal{I}_{\mathcal{X}_i}(X_t)$ by the mean $N^{-1}
\sum_{j=1}^N \mathcal{I}_{\mathcal{X}_i}(X_t^{(j)})$, that is the proportion of
chains currently in set $\mathcal{X}_i$ (Algorithm 2 line
\ref{algo:generalizedWL:nu}). Likewise the update of the bias uses all the
chains, again replacing the indicator function by the proportion of chains
currently in a given set (Algorithm 2 line \ref{algo:generalizedWL:bias}).  We
have therefore replaced indicator functions in the Wang-Landau algorithm by the
law of the MCMC chain associated with the current parameter. Since this law is
not accessible, we perform a mean field approximation at each time step.  A
similar expression has recently been employed by \citet{Liang2011a} for use in
parallelizing the stochastic approximation Monte Carlo algorithm
\citep{Liang2007a, Liang2009a}.  Note that while we have designed the chains to
communicate at each iteration, such frequent message passing can be costly,
particularly on graphics processing units.  In such situations, one could alter
the algorithm such that the chains only communicate periodically.

Our results (see Section \ref{sec:applications}) show that $N$ interacting
chains run for $T$ iterations can strongly outperform a single chain run for $N
\times T$ iterations, in terms of variance of the resulting estimates.
Specifically, having a sample approximately distributed according to
$\pi_{\theta_{t}}(x)$ instead of a single point at iteration $t$ improves and
stabilizes the subsequent estimate $\theta_{t+1}$.  We explore the tradeoff
between $N$ and $T$ in more detail in Section \ref{sec:applications}.

Note that, while the original single-chain Wang-Landau algorithm was not
straightforward to parallelize due to its iterative nature, the proposed
algorithm can strongly benefit from multiple processing units: at a given
iteration the $N$ move steps can be done in parallel, as long as the results
are consequently collected to update the bias before the next iteration.
Therefore if multiple processors are available, as e.g. in recent central
processing units and in graphics processing units (see e.g. \citet{Lee2009a,
Suchard2009a}), the computational cost can be reduced much more than what was
possible with the single-chain Wang-Landau algorithm.  To summarize, the
proposed use of interacting chains can both improve the convergence of the
estimates, regardless of the number of available processors, and additionally
benefit from multiple processors.

Finally, an additional benefit of using $N$ parallel chains is that they can
start from various points, drawn from the initial distribution $\pi_0$; hence
if $\pi_0$ is flat enough, the chains can start from different local modes,
which improves \emph{de facto} the exploration. However, we show in Section
\ref{sec:applications} that the chains still explore the space even if they
start within the same mode, and hence the efficiency of the method does not
rely on the choice of $\pi_0$, contrary to what we observed with sequential
Monte Carlo methods.  Additionally, because the sampler is attempting to
explore both the state-space as well as the range of the reaction coordinate
simultaneously, our parallel formulation allows the sampler to borrow strength
between chains, providing for exploration of the reaction coordinate without
having to move a single chain across potentially large and high-dimensional
state-spaces to traverse the reaction coordinate values.

\subsection{Adaptive Proposal Mechanism \label{sec:proposals}}

As discussed earlier, it is important to automate the proposal mechanism to
improve movement across the state space.  A well-studied proxy for optimal
movement is the algorithm's Metropolis-Hastings acceptance rate.  Too low an
acceptance rate signifies the algorithm is attempting to make moves that are
too large, and are therefore rejected.  Too high an acceptance rate signifies
the algorithm is barely moving.  As such, we suggest adaptively tuning the
proposal variance to encourage an acceptance rate of $0.234$ as recommended in
\citet{Roberts1997a}, although we have found settings in the range $0.1$ to
$0.5$ to work well in all examples tested.  The Robbins-Monro stochastic
approximation update of the proposal standard deviation $\sigma_t$ is as
follows:
\begin{align}
	\sigma_{t+1} = \sigma_t + \rho_t \left( 2 \mathcal{I}(A > 0.234) - 1 \right)
\label{robbinsMonroeProposal}
\end{align}
where $t$ is the current iteration of the algorithm, $\rho_t$ is a decreasing
sequence (typically $\rho_t = 1 / t$), and $A$ is the acceptance rate
(proportion of accepted moves) of the particles. Through this update, the
proposal variance grows after samples are accepted, and shrinks when samples
are rejected, encouraging exploration of the state space.

Another approach to adaptively tuning the proposal distribution is to use the
following mixture of Gaussian random-walks:
\begin{align}
	X^* \sim w_1 \mathcal{N}\left({X}_{t-1},
\frac{(2.38)^2}{p}\Sigma_t \right)
+ w_2 \mathcal{N}\left({X}_{t-1},
\frac{(\sigma_I)^2}{p}I_{p}\right)
\label{mixtureProposal}
\end{align}
with $w_1 + w_2 = 1$, $\Sigma_t$ the empirical covariance of the chain history
-- an estimator of the covariance structure of the target -- and $I_{p}$ the $p
\times p$ identity matrix where $p$ is the dimension of the target space. The
first component of this mixture makes the proposal adaptive and able to learn
from the past, while the second component helps to explore the space. For
instance, if the chain is stuck in a mode, the first component's variance might
become small, yet the second component guarantees a chance to eventually escape
the mode. Hence the second component acts as a ``safety net'' and therefore its
weight is small, typically $w_2 = 0.05$, and its standard deviation $\sigma_I$
may be set large to improve mixing \citep{guan2007small}.

In our context where parallel chains are run together, we use all the chains to
estimate the empirical covariance $\Sigma_t$ at each iteration. Note that the
computation of this covariance does not require the storage of the whole
history of the chain and can be done at constant cost, since recurrence
formulae exist to compute the empirical covariance, as explained for instance
in \citet{Welford1962a}.  The value $2.38^2$ is justified by asymptotic
optimality reflections on certain classes of models (see, e.g.,
\citet{Roberts1997a} and \citet{Roberts2009a}).

\subsection{Using the Output to Perform Inference}

While the resulting samples from the proposed algorithm PAWL are not from
$\pi$, but rather an approximation of the biased version (\ref{BiasedDensity}),
one can use importance sampling or advanced sequential Monte Carlo ideas to
transition the samples to $\pi$ (see \citet{Chopin2010b} for details).
Alternatively, the samples from the exploratory algorithm can be used to seed a
more traditional MCMC algorithm, as advocated by \citet{Schmidler2011a}.

The pseudo-code for PAWL, combining the parallel Wang-Landau algorithm with
adaptive binning and proposal mechanisms, is given in the Appendix.  Before
proceeding to examples, it is important to reiterate the importance of the
values $\psi(i) = \int_{\mathcal{X}_i} \pi(x) \mathrm{d}x$.  Specifically,
certain choices of the reaction coordinate $\xi{(x)}$ result in $\psi(i)$
having inherent value.  For example, it is possible in a model selection
application to use the model order as $\xi{(x)}$, in which case the values
$\psi(i)$ could be employed to calculate Bayes factors and other quantities of
interest.

\section{Applications\label{sec:applications}}

\label{section:ref}

We now demonstrate PAWL applied to three examples including variable selection,
mixture modeling, and spatial imaging.  A fourth pedagogical example is
available in the appendices.  In each application we walk through our
proposed algorithm (described explicitly as Algorithm \ref{algo:final} in the
Appendix), first running preliminary (adaptive) Metropolis-Hastings MCMC to
determine the initial range for the reaction coordinate $\xi$ and initial
values for the proposal parameters and starting state of the interacting
Wang-Landau chains.  This range is then increased to encourage exploration of
low-density regions of the space, and an initial number of bins is specified.
Once this groundwork is set, the same Metropolis-Hastings algorithm run in the
preliminary stage is embedded within the PAWL algorithm.

\subsection{\texorpdfstring{$g$-Prior Variable Selection}{g-Prior Variable Selection}}

We proceed by conducting variable selection on the pollution data set of
\citet{McDonald1973a}, wherein mortality is related to pollution levels through
$15$ independent variables including mean annual precipitation, population per
household, and average annual relative humidity.  Measured across 60
metropolitan areas, the response variable $\mathbf{y}$ is the age-adjusted
mortality rate in the given metropolitan area.  Our goal is to identify the
pollution-related independent variables which best predict the response.  With
15 variables, calculating the posterior probabilities of the $32,768$ models
exactly is possible but time-consuming.  We have chosen this size of data set
to provide for difficult posterior exploration, yet allow a study of
convergence of $\theta$ towards $\psi$.

With an eye towards model selection, we introduce the binary indicator variable
$\gamma \in \{0,1\}^p$, where $\gamma_j=1$ means the variable
$\mbox{\boldmath$x$}_j$ is included in the model. As such, $\gamma$ can
describe all of the $2^p$ possible models.  Consider the normal likelihood
\begin{align*}
\tag{5}\mathbf{y}|\mu,\mathbf{X},\mbox{\boldmath$\beta$},\sigma^2 \sim N_n(\mathbf{X}\mbox{\boldmath$\beta$},\sigma^2 \mathbf{I}_n).
\end{align*} 

If $\mathbf{X}_\gamma$ is the model matrix which excludes all $\mbox{\boldmath$x$}_j$'s if $\gamma_j = 0$, we can employ
the following prior distributions for $\mbox{\boldmath$\beta$}$ and $\sigma^2$ \citep{Zellner1986a, Marin2007a}:
\begin{align*}
\pi(\bm{\beta}_\gamma,\sigma^2 | \gamma) \propto (\sigma^2)^{-(q_\gamma +
1)/2 - 1} \exp \left[ -\frac{1}{2g\sigma^2}  \mbox{\boldmath$\beta$}_\gamma^T
(\mathbf{X}_\gamma^T\mathbf{X}_\gamma)\mbox{\boldmath$\beta$}_\gamma \right].
\end{align*}
where $q_{\gamma} = \mbox{\textbf{1}}_n^T\gamma$ represents the number of
variables in the model.  While selecting $g$ can be a difficult problem, we
have chosen it to be very large ($g = \exp(20)$) to induce a sparse model,
which is difficult to explore due to the small marginal probabilities of most
variables.  After integrating over the regression coefficients $\bm{\beta}$,
the posterior density for $\gamma$ is thus
\begin{align*}
\tag{6} \pi(\gamma | \mathbf{y}, \mathbf{X}) \propto (g+1)^{-(q_\gamma + 1)/2} \left[
\mathbf{y}^T\mathbf{y} - \frac{g}{g+1}\mathbf{y}^T\mathbf{X}_\gamma (\mathbf{X}_\gamma^T
\mathbf{X}_\gamma)^{-1}\mathbf{X}_\gamma\mathbf{y} \right]^{-n/2}.
\end{align*}

While we select the log energy function $-\log \pi(x)$ as the reaction
coordinate $\xi(x)$ for our analysis, it is worth noting that many other
options exist.  For instance, it would be natural to consider the model
saturation $q_{\gamma}/p$, which would ensure exploration across the different
model sizes.  However, we select $\xi(x) = -\log \pi(x)$ to emphasize the
universality of using the energy function as the reaction coordinate.  

We first run a preliminary Metropolis-Hastings algorithm which flips a variable
on/off at random, accepting or rejecting the flip based on the resulting
posterior densities.  Due to high correlation between variables, a better
strategy might be to flip multiple variables at once; however, we restrain from
exploring this to demonstrate PAWL's ability to make viable even poorly
designed Metropolis-Hastings algorithms.  The preliminary algorithm run found
values $377 < -\log \pi(x) < 410$, which we extend slightly to create $20$
equally spaced bins in the range $[377, 450]$.  It is worth reiterating that
the resulting samples generated from PAWL are from a biased version of $\pi$;
as such, importance sampling techniques could be used to recover $\pi$, or the
samples obtained could be used to seed a more traditional MCMC algorithm.

\begin{figure}
    \centering
    \includegraphics[width=\textwidth]{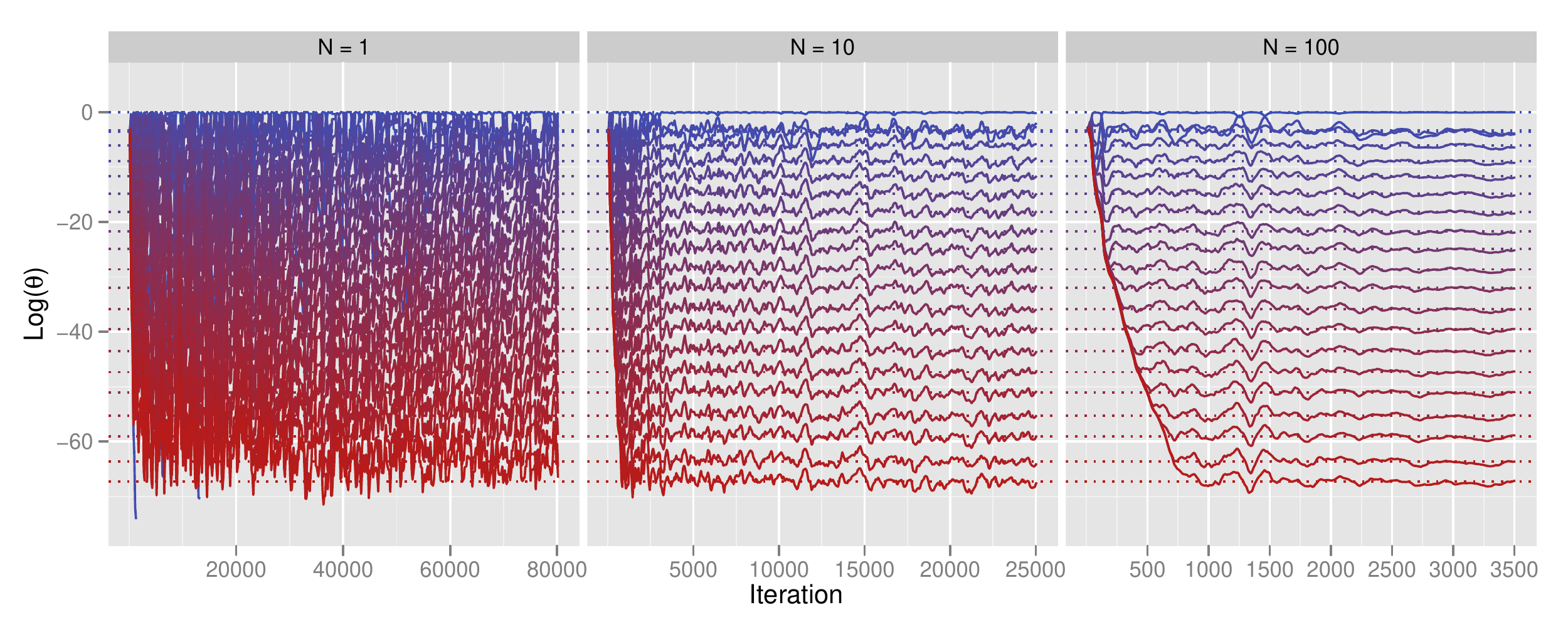}
    \caption{Variable selection example: convergence of Wang-Landau for $N=1, 10,
    100$. Iterations set such that each algorithm runs in $2$ minutes ($\pm 5$
    seconds). $\theta$ for each bin shown in solid lines.  True values ($\psi$)
    shown as dotted lines.}
    \label{gPriorConvergence}
\end{figure}

\begin{figure}
    \centering
    \includegraphics[width=\textwidth]{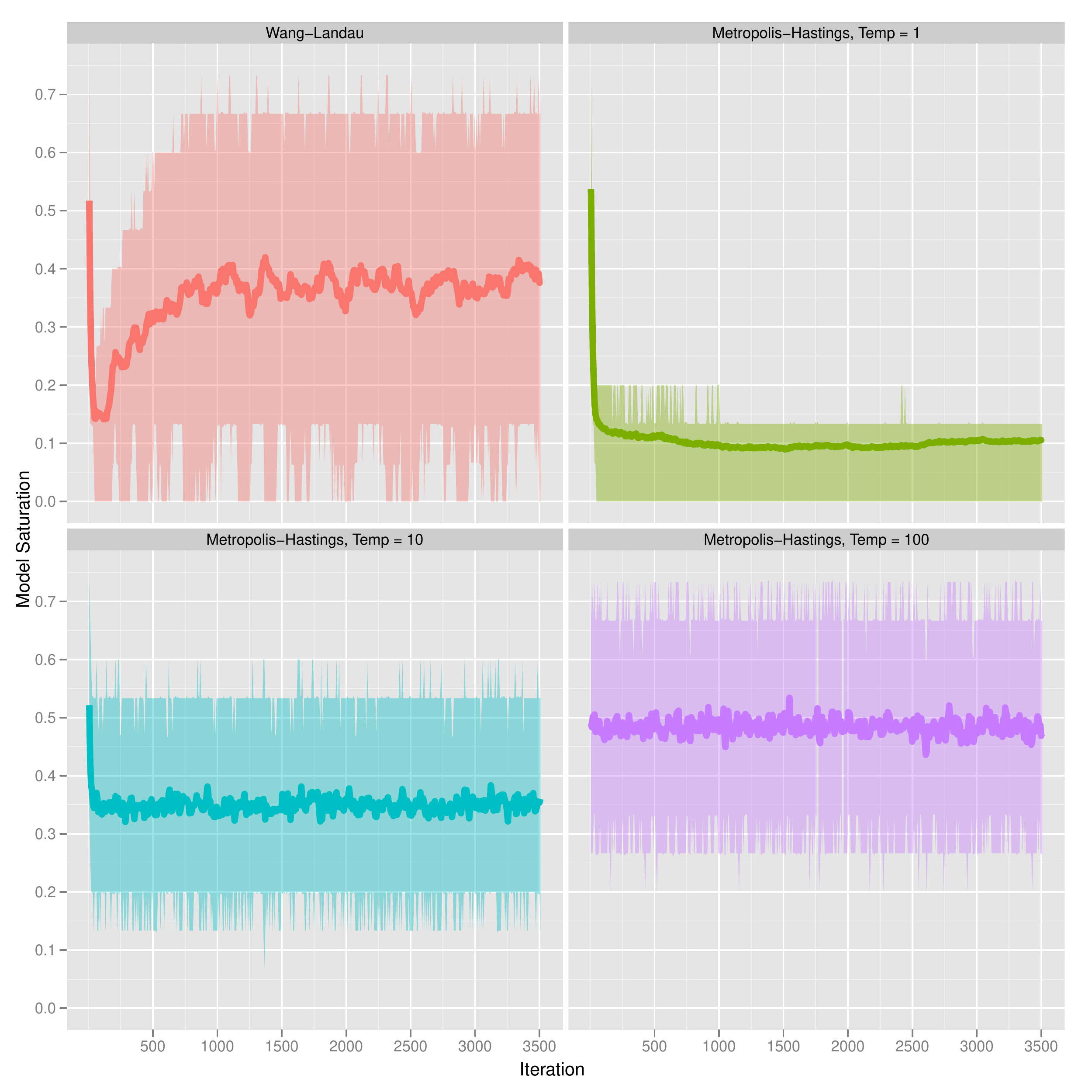}
    \caption{Variable selection example: exploration of model space by PAWL
    and Metropolis-Hastings at $3$ temperature settings. (a) Model Saturation
    (proportion of non-zero variables in model) as function of algorithm iterations.
    The solid lines are the mean of $N=100$ chains, while the shaded regions are
    the middle $95\%$ of the chains.}
    \label{gPriorExploration}
\end{figure}

Due to the size of the problem, we are able to enumerate all posterior values,
and hence may calculate $\psi$ exactly.  As such, we begin by examining the
effect of the number of particles $N$ on the parallel Wang-Landau algorithm.
To further focus on this aspect, we suppress adaptive binning and proposals for
this example.  Figure \ref{gPriorConvergence} shows the convergence of $\theta$
to $\Psi$ for $N=1,10, 100$.  We see that the algorithm's convergence improves
with more particles.  Using $N=100$ particles, we now examine PAWL compared to
the Metropolis-Hastings algorithm (run on $N$ chains) mentioned above on the
unnormalized targets $\pi, \pi^{1/10}, \pi^{1/100}$.  Consider Figure
\ref{gPriorExploration}; on the target distribution $\pi$, the
Metropolis-Hastings algorithm becomes stuck in high-probability regions.
However, on the tempered distributions, the algorithm explores the space more
thoroughly, although not to the same level as PAWL.  Specifically, PAWL
explores a much wider range of models, including the highest probability
models, whereas the tempered distributions do not.  Here the Wang-Landau
algorithm as well as the Metropolis-Hastings algorithm both use $N=100$ chains
for $T=3500$ iterations, the former taking $253 \pm 13$ seconds and the latter
taking $247 \pm 15$ seconds across 10 runs, indicating that the additional cost
for PAWL is negligible.

\subsection{Mixture Modeling}

Mixture models provide a challenging case of multimodality, due partly to the
complexity of the model and partly to a phenomenon called ``label switching''
(see e.g. \citet{FruhwirthMixture} for a book covering Bayesian inference for
these models, \citet{Diebolt:Robert:1994} and \citet{Richardson:Green:1997} for
seminal papers using MCMC for mixture models, and \citet{stephens:2000b} and
\citet{Jasra:2005} on the label switching problem).  Articles describing
explorative MCMC algorithms often take these models as benchmarks, as e.g.
population MCMC and SMC methods in \citet{Jasra2007b}, the Wang-Landau
algorithm in \citet{Atchade2010a}, free energy methods in
\citet{ChopinLelievreStoltz,Chopin2010b}, and parallel tempering with
equi-energy moves in \cite{Baragatti2012a}. 

Consider a Bayesian Gaussian mixture model, i.e. for $i=1,\ldots n$,
\[
p(y_{i}|q,\mu,\lambda)=\sum_{k=1}^{K}q_{k}\,\varphi(y_{i};\mu_{k},\lambda_{k}^{-1}),\]
where $\varphi$ is the probability density function of the Gaussian distribution, 
$K$ is the number of components, $q_k$, $\mu_k$ and $\lambda_k$ are respectively the weight, 
the mean and the precision of component $k$. The component index $k$ is also called its label. 
Following \citet{Richardson:Green:1997}, the prior is taken as, for $k=1,\ldots,K$,
\begin{align*}
\mu_{k} & \sim  \mathrm{N}(M,\kappa^{-1}),&
\lambda_{k} & \sim  \mathrm{Gamma}(\alpha,\beta),\\
\beta & \sim  \mathrm{Gamma}(g,h),&
(q_{1},\ldots,q_{K-1}) & \sim  \mathrm{Dirichlet}_{K}(1,\ldots,1)
\end{align*}
with, e.g., $\kappa=4/R^{2}$, $\alpha=2$, $g=0.2$, $h=100g/\alpha R^{2}$,
$M=\bar{y},$ $R=$range$(y)$.

The invariance of the likelihood to permutations in the labelling of the
components leads to the ``label switching'' problem: since there are $K!$
possible permutations of the labels, each mode has necessarily $K! - 1$
replicates. 
We emphasize that this model has been thoroughly studied and is hence
well-understood from a modeling point of view, but it still induces a
computationally challenging sampling problem for which difficulty can be
artificially increased through the number of components $K$.

Note that in this parametrization $\beta$, the rate of the Gamma prior
distribution of the precisions $\lambda_k$, is estimated along with the
parameters of interest $q_{1:K-1}$, $\mu_{1:K}$ and $\lambda_{1:K}$.
\citet{ChopinLelievreStoltz,Chopin2010b} suggest that $\beta$ can be used as a
reaction coordinate, since a large value of $\beta$ results in a small
precision and hence in a flatter posterior distribution of the other
parameters, which is easier to explore than the distribution associated with
smaller values of $\beta$; we refer to these articles for further exploration
of this choice of reaction coordinate, and instead default to $\xi(x) = - \log
\pi(x)$.

We create a synthetic $100$-sample from a Gaussian mixture with $k = 4$
components, weights $1/4$, means $-3$, $0$, $3$, $6$ and variances $0.55^{2}$
as in \citet{Jasra:2005}.  The goal is to explore the highly multimodal
posterior distribution of the $13$-dimensional parameter $\theta =
(w_{1:4},\mu_{1:4}, \lambda_{1:4}, \beta)$ where $w_k$ is the unnormalized
weight: $q_k = w_k / \sum_{k=1}^K w_k$.  Unnormalized weights may be handled
straightforwardly in MCMC algorithms since they are defined on $\mathbb{R}^+$
and not on the $K$-simplex as with the $q_k$. 

The proposed algorithm is compared to a Sequential Monte Carlo sampler (SMC)
and a parallel adaptive Metropolis--Hastings (PAMH) algorithm, that we detail
below.  We admittedly use naive versions of these competitors, arguing that
most improvements of these could be carried over to PAWL.  For instance, a
mixture of Markov kernels as suggested for the SMC algorithm in Section 3.2 of
\citet{Jasra2007b} can be used in the proposal distribution of PAWL; and since
PAWL is a population MCMC algorithm, exchange and crossover moves could be used
as well, as suggested for the Population SAMC algorithm in \cite{Liang2011a}.
To get a plausible range of values for the reaction coordinate of the proposed
algorithm without user input, an initial adaptive MCMC algorithm is run with $N
= 10$ chains and $T^{\text{init}} = 1,000$ iterations. The initial points of
these chains are drawn from the prior distribution of the parameters. This
provides a range of log density values, from which we compute the $10\%$ and
$90\%$ empirical quantiles, denoted by $q_{10}$ and $q_{90}$ respectively.  In
a conservative spirit, the bins are chosen to equally divide the interval
$[q_{10}, \;q_{10} + 2 (q_{90} - q_{10})]$ in $20$ subsets.  Hence the
algorithm is going to explore log density values in a range that is
approximately twice as large as the values initially explored.  Note that we
use quantiles instead of minimum and maximum values to make the method more
robust.

Next, PAWL itself is run for $T = 200,000$ iterations, starting from the
terminal points of the $N$ preliminary chains, resulting in a total number of
$N (T + T_{\text{init}})$ target density evaluations.  In this situation, even
with only $100$ data points, most of the computational cost goes into the
evaluation of the target density. This confirms that algorithmic parameters
such as the number of bins does not significantly affect the overall
computational cost, at least as long as the target density is not extremely
cheap to evaluate. The adaptive proposal is such that it targets an acceptance
rate of $23.4\%$.  Meanwhile the PAMH algorithm using the same adaptive
proposal is run with $N = 10$ chains and $T^\star = 250,000$ iterations, hence
relying on more target density evaluations for a comparable computational cost.

Finally, the SMC algorithm is run on a sequence of tempered distribution
$(\pi_k)_{k=1}^{K}$, each density being defined by: \[\pi_k(x) \propto
\pi^{\zeta_k}(x) p_0^{1 - \zeta_k}(x)\] where $p_0$ is an initial distribution
(here taken to be the prior distribution), and $\zeta_k = k / K$. The number of
steps $K$ is set to $100$ and the number of particles to $40,000$. When the
Effective Sample Size (ESS) goes below $90\%$, we perform a systematic
resampling and $5$ consecutive Metropolis--Hastings moves. We use a random walk
proposal distribution, which variance is taken to be $ c \hat\Sigma$ where
$\hat\Sigma$ is the empirical covariance of the particles and $c$ is set to
$10\%$; see \cite{Jasra2007b} for more details. The parameters are chosen to
induce a computational cost comparable to the other methods. However for the
SMC sampler the number of target density evaluations is a random number, since
it depends on the random number of resampling steps: the computational cost is
in general less predictable than using MCMC.

\begin{figure}
    \centering
    \subfigure[Locations of the global modes of the posterior 
    distribution projected on $(\mu_1, \mu_2)$]{
    \label{subfig:mixture:targets}\includegraphics[width=0.42\textwidth]{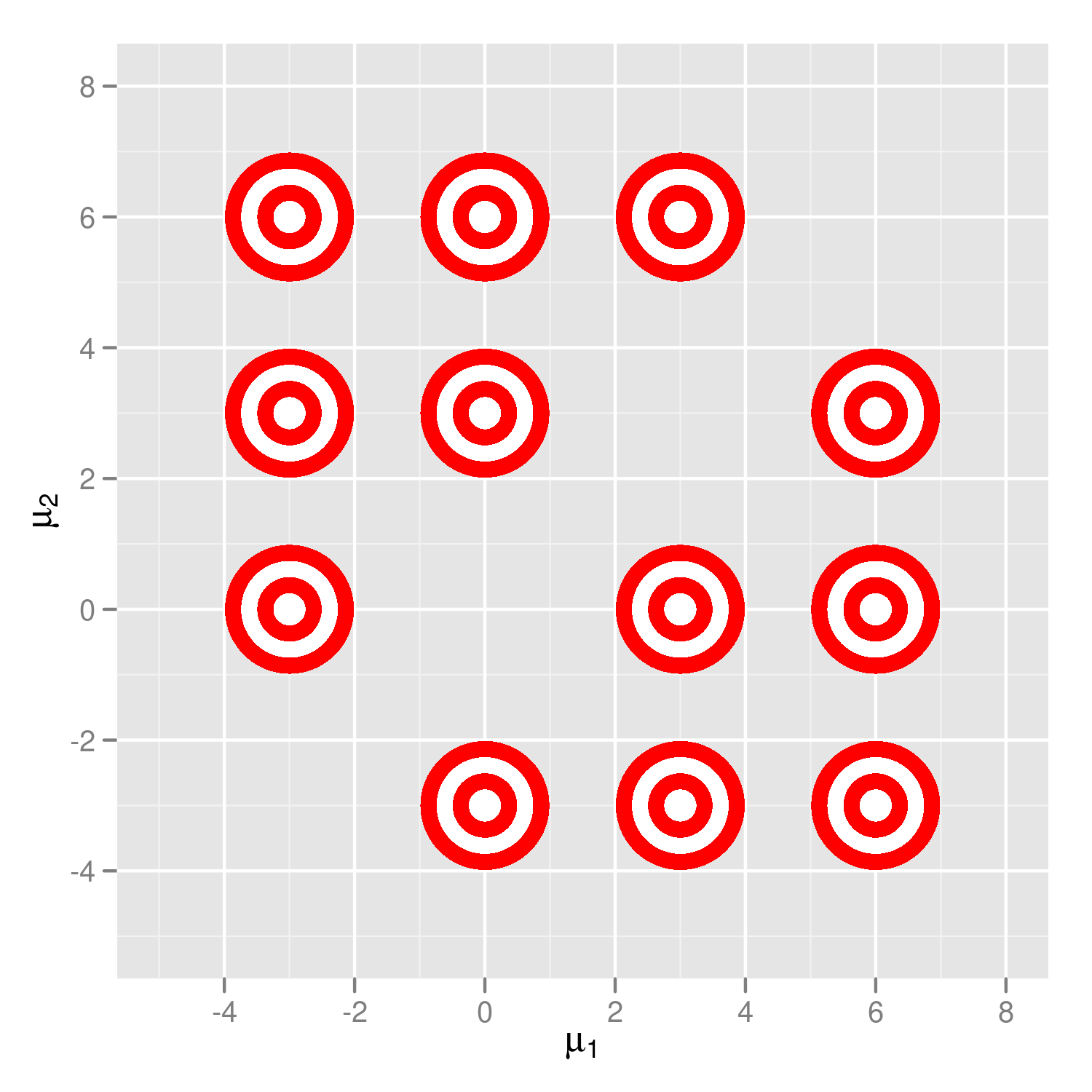}}
    \subfigure[Projection of the chains generated by the 
    parallel adaptive MH algorithm on $(\mu_1, \mu_2)$]{
    \label{subfig:mixture:amcmc}\includegraphics[width=0.42\textwidth]{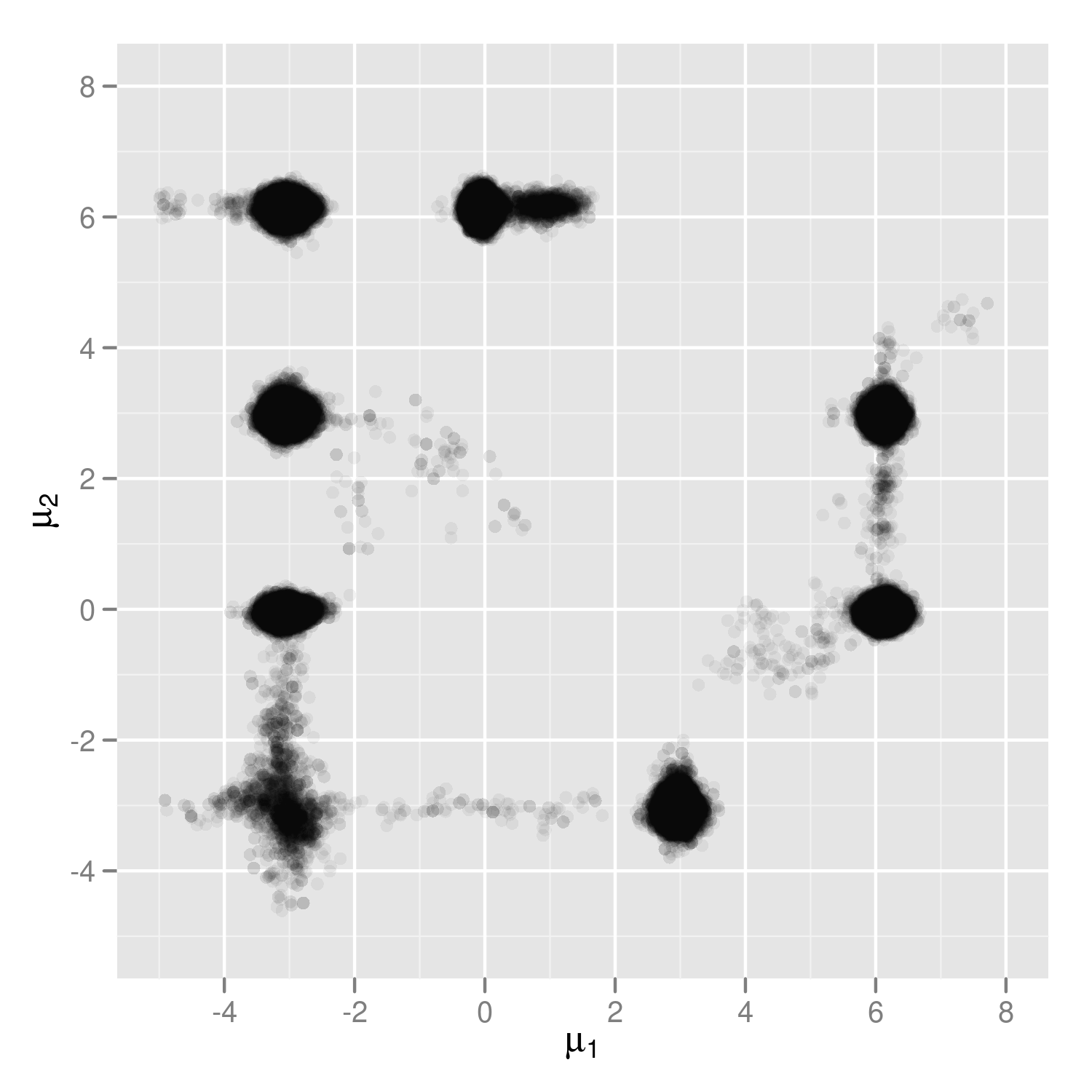}}
    \subfigure[Projection of the chains generated by PAWL on $(\mu_1, \mu_2)$]{
    \label{subfig:mixture:pwl}\includegraphics[width=0.42\textwidth]{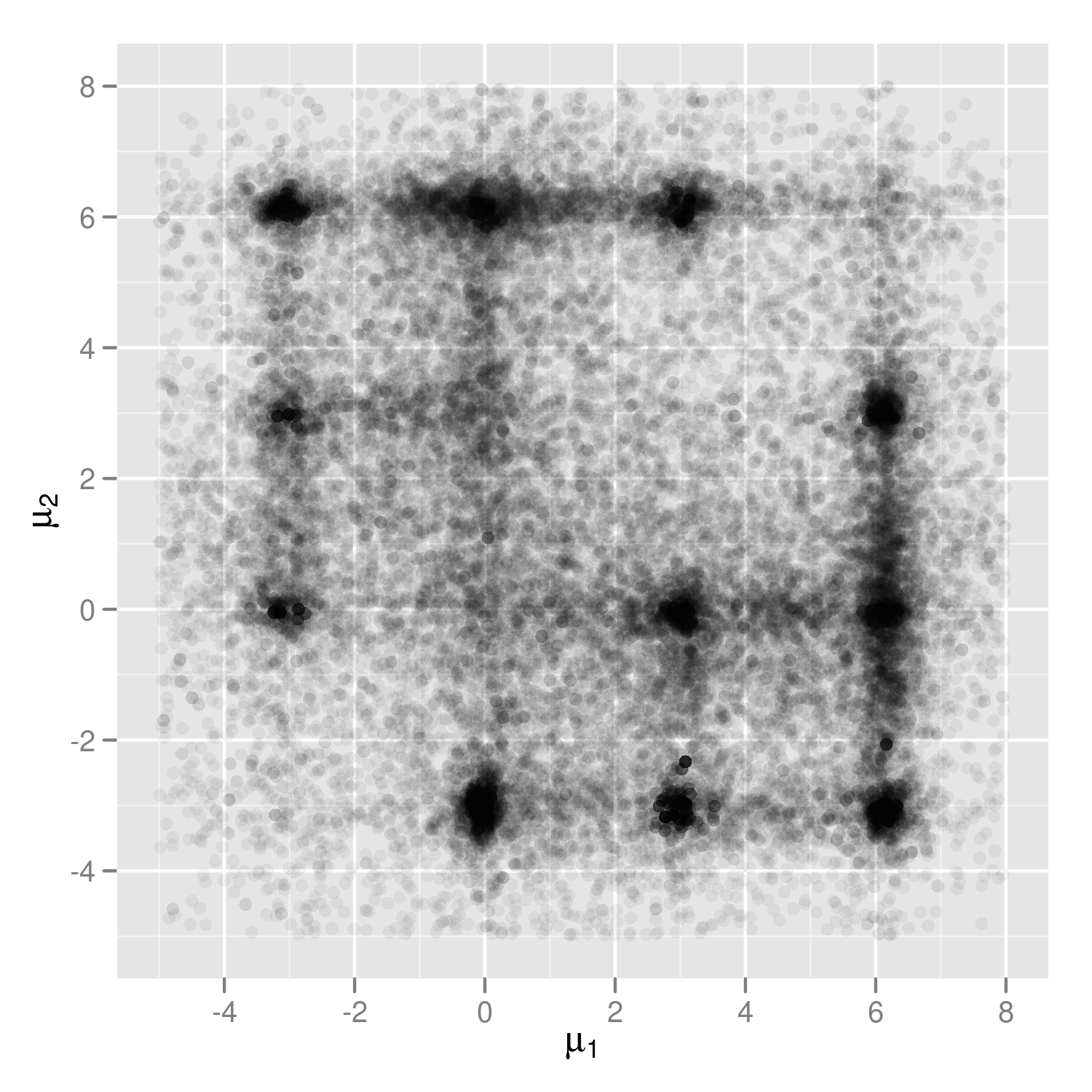}}
    \subfigure[Projection of the particles generated by the 
    SMC algorithm on $(\mu_1, \mu_2)$]{
    \label{subfig:mixture:smc}\includegraphics[width=0.42\textwidth]{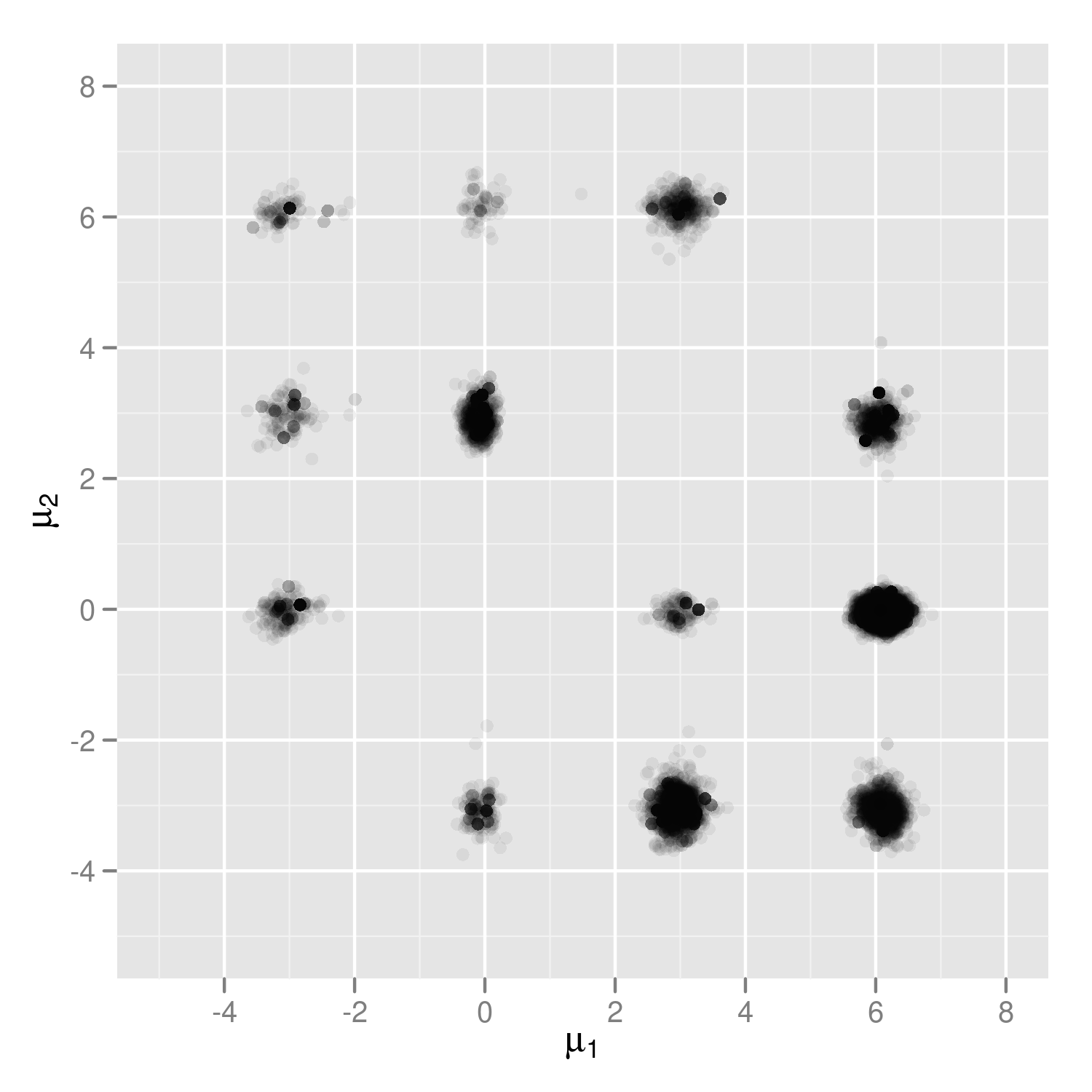}}
    \caption{
    \label{fig:mixture:projectionMuMu}Mixture model example: exploration of
    the posterior distribution projected on the $\mu_1, \mu_2$ plane, using different algorithms.}
    \label{mixture}
\end{figure}

First we look at graphical representations of the generated samples.  Figure
\ref{fig:mixture:projectionMuMu} shows the resulting points projected on the
$(\mu_1,\mu_2)$ plane, restricted on $[-5, 9]^2$. In this plane there are $12$
replicates of each mode, indicated by target symbols in Figure
\ref{subfig:mixture:targets}. These projections do not allow one to check that
all the modes were visited since they project the $13$-dimensional target space
on a $2$-dimensional space.  Figure \ref{subfig:mixture:amcmc} shows that the
adaptive MCMC method clearly misses some of the modes, while visiting many
others.  Figure \ref{subfig:mixture:pwl} shows how the chains generated by the
modified Wang-Landau algorithm easily explore the space of interest, visiting
both global and local modes.  To recover the main modes from these chains, we
use the final value of the bias, $\theta_T$, as importance weights to correct
for the bias induced by the algorithm; in Figure \ref{subfig:mixture:pwl} the
importance weights define the transparency of each point: the darker the point,
the more weight it has. Finally Figure \ref{subfig:mixture:smc} shows how the
SMC sampler also put particles in each mode; again the transparency of the
points is proportional to their weights.

We now turn to more quantitative measures of the error made on marginal
quantities. Since the component means admit $4$ identical modes around $-3$,
$0$, $3$ and $6$, we know that their overall mean is approximately equal to
$\mu^\star = 1.5$. We then compute the following error measurement:
\[\text{error} = \sqrt{\sum_{k=1}^K \left(\hat{\mu}_k - \mu^\star\right)^2} \]
where $\hat{\mu}_k$ is the mean of the generated sample (taking the weights
into account for PAWL and SMC).  Table \ref{table:mixture:goodinit} shows the
results averaged over $10$ independent runs: the means of each component, the
error defined above and the (wall-clock) run times obtained using the same CPU,
along with their standard deviations.  The results highlight that in this
context PAWL gives more precise results than SMC, for the same or less
computational cost; the comparison between parallel MCMC and SMC confirms the
results obtained in \cite{Jasra2007b}. The small benefit of PAWL over PAMH can
be explained by considering the symmetry of the posterior distribution: even if
some modes are missed by PAMH as shown in Figure \ref{subfig:mixture:amcmc},
the approximation of the posterior expectation might be accurate, though the
corresponding variance will be higher.

\begin{table}[ht]
    \begin{center}
        \begin{tabular}{|l|l|l|l|l|l|l|}
            \hline
            Method & $\mu_1$ & $\mu_2$ & $\mu_3$ & $\mu_4$ & Error & Time (s) \\ 
            \hline
            PAWL & 1.42 $\pm$ 0.99 & 1.42 $\pm$ 0.58 & 1.39 $\pm$ 0.90 & 1.75 $\pm$ 0.78 & 1.50 $\pm$ 0.59 & 209 $\pm$ 1 \\ 
            PAMH & 1.58 $\pm$ 0.81 & 1.25 $\pm$ 0.72 & 1.04 $\pm$ 1.07 & 2.09 $\pm$ 1.00 & 1.75 $\pm$ 0.80 & 233 $\pm$ 1 \\ 
            SMC  & 1.00 $\pm$ 1.96 & 2.99 $\pm$ 1.38 & 0.92 $\pm$ 2.27 & 1.10 $\pm$ 2.11 & 3.89 $\pm$ 1.34 & 269 $\pm$ 7 \\ 
            \hline
        \end{tabular}
    \end{center}
    \caption{\label{table:mixture:goodinit} Estimation of the means of the mixture components,
    for the proposed method (PAWL), Parallel Adaptive Metropolis--Hastings (PAMH) and Sequential Monte Carlo (SMC),
    using the prior as initial distribution.
    Quantities averaged over $10$ independent runs for each method.}
\end{table}

Next we consider a more realistic setting, where the initial distribution is
not well spread over the parameter values: instead of taking the prior
distribution itself, we use a similar distribution but with an hyperparameter
$\kappa$ equal to $1$ instead of $4/R^2$, which for our simulated data set is
equal to $0.03$. This higher precision makes the initial distribution
concentrated on a few modes, instead of being fairly flat over the whole region
of interest. We keep the prior unchanged, so that the posterior is left
unchanged. For PAWL and PAMH, this means that the initial points of the chains
are all close one to another; and likewise for the initial particles in the SMC
sampler.  The results are shown in Table \ref{table:mixture:badinit}, and
illustrate the degeneracy of SMC when the initial distribution is not
well-chosen; though this is not surprising, this is important in terms of
exploratory algorithms when one does not have prior knowledge of the region of
interest. Both parallel MCMC methods give similar results as with the previous,
flatter initial distribution.

\begin{table}[ht]
    \begin{center}
        \begin{tabular}{|l|l|l|l|l|l|l|}
            \hline
            Method & $\mu_1$ & $\mu_2$ & $\mu_3$ & $\mu_4$ & Error & Time \\ 
            \hline
            PAWL & 1.16 $\pm$ 0.75 & 2.04 $\pm$ 0.50 & 1.72 $\pm$ 0.80 & 1.07 $\pm$ 1.22 & 1.48 $\pm$ 1.10 & 210 $\pm$ 1 \\ 
            PAMH & 1.37 $\pm$ 0.73 & 1.48 $\pm$ 1.39 & 1.71 $\pm$ 0.81 & 1.44 $\pm$ 1.11 & 1.75 $\pm$ 1.01 & 234 $\pm$ 1 \\ 
            SMC  & 0.35 $\pm$ 2.13 & 0.82 $\pm$ 1.55 & 3.19 $\pm$ 2.41 & 1.62 $\pm$ 1.85 & 4.17 $\pm$ 1.41 & 337 $\pm$ 8 \\ 
            \hline
        \end{tabular}
    \end{center}
    \caption{\label{table:mixture:badinit} Estimation of the means of the mixture components,
    for the proposed method (PAWL), Parallel Adaptive Metropolis--Hastings (PAMH) and Sequential Monte Carlo (SMC),
    using a concentrated initial distribution.
    Quantities averaged over $10$ independent runs for each method.}
\end{table}

Finally, we compare different algorithmic settings for the PAWL algorithm,
changing the number of chains and the number of iterations.  The results are
shown in Table \ref{table:mixture:differentN}.  First we see that, even on a
single CPU, the computing time is not exactly proportional to $N\times T$, the
number of target density evaluation.  Indeed the computations are vectorized by
iteration, and hence it is typically cheaper to compute one iteration of $N$
chains than $N$ iterations of $1$ chain; although this would not hold for every
model. We also see that the algorithm using only one chain failed to explore
the modes, resulting in a huge final error. Finally we see that with $50$
chains and only $50,000$ iterations, the algorithm provides results of
approximately the same precision as with $10$ chains and $200,000$ iterations.
This suggests that the algorithm might be particularly interesting if parallel
processing units are available, since the computational cost would then be much
reduced.

\begin{table}[ht]
    \begin{center}
        \begin{tabular}{|l|l|l|l|l|l|l|}
            \hline
            Parameters & $\mu_1$ & $\mu_2$ & $\mu_3$ & $\mu_4$ & Error & Time \\ 
            \hline
            $N = 1$ &  
            \multirow{2}{*}{0.37 $\pm$ 3.46}  & \multirow{2}{*}{2.01 $\pm$ 3.27} & 
            \multirow{2}{*}{2.53 $\pm$ 3.04}  & \multirow{2}{*}{0.95 $\pm$ 3.46} &
            \multirow{2}{*}{6.39 $\pm$ 1.30}  & \multirow{2}{*}{265 $\pm$ 40} \\ 
            $T = 5\times10^5$ & & & & & & \\
            \hline
            $N = 10$ & 
            \multirow{2}{*}{1.42 $\pm$ 0.99}  & \multirow{2}{*}{1.42 $\pm$ 0.58} & 
            \multirow{2}{*}{1.39 $\pm$ 0.90}  & \multirow{2}{*}{1.75 $\pm$ 0.78} &
            \multirow{2}{*}{1.50 $\pm$ 0.59}   & \multirow{2}{*}{209 $\pm$ 1} \\ 
            $T = 2\times10^5$ & & & & & & \\
            \hline
            $N = 50$ & 
            \multirow{2}{*}{1.51 $\pm$ 0.88}  & \multirow{2}{*}{1.5 $\pm$ 0.9} & 
            \multirow{2}{*}{1.65 $\pm$ 0.64}  & \multirow{2}{*}{1.31 $\pm$ 0.31} &
            \multirow{2}{*}{1.22 $\pm$ 0.69}  & \multirow{2}{*}{178 $\pm$ 2} \\ 
            $T = 5\times10^4$ & & & & & & \\
            \hline
        \end{tabular}
    \end{center}
    \caption{\label{table:mixture:differentN} Estimation of the means of the mixture components,
    for the proposed method (PAWL), for different values of $N$, the number of chains, and $T$,
    the number of iterations.
    Quantities averaged over $10$ independent runs for each set of parameters.}
\end{table}

\subsection{Spatial Imaging}

\begin{figure}
    \centering
    \subfigure[Original Image]{\label{subfig:IsingModelAll}\includegraphics[width=0.46\textwidth]{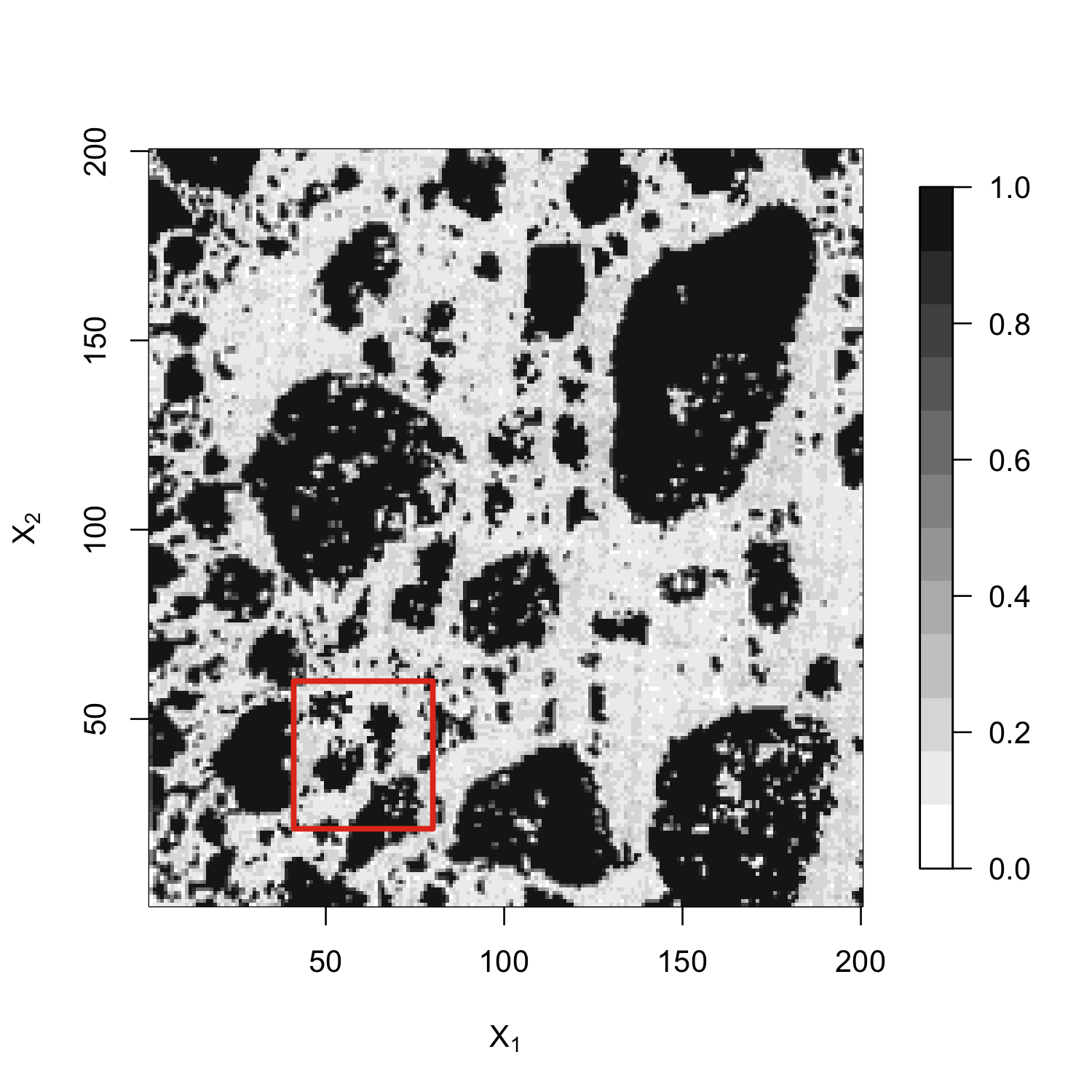}}
    \subfigure[Focused Region of Image]{\label{subfig:IsingModelZoom}\includegraphics[width=0.46\textwidth]{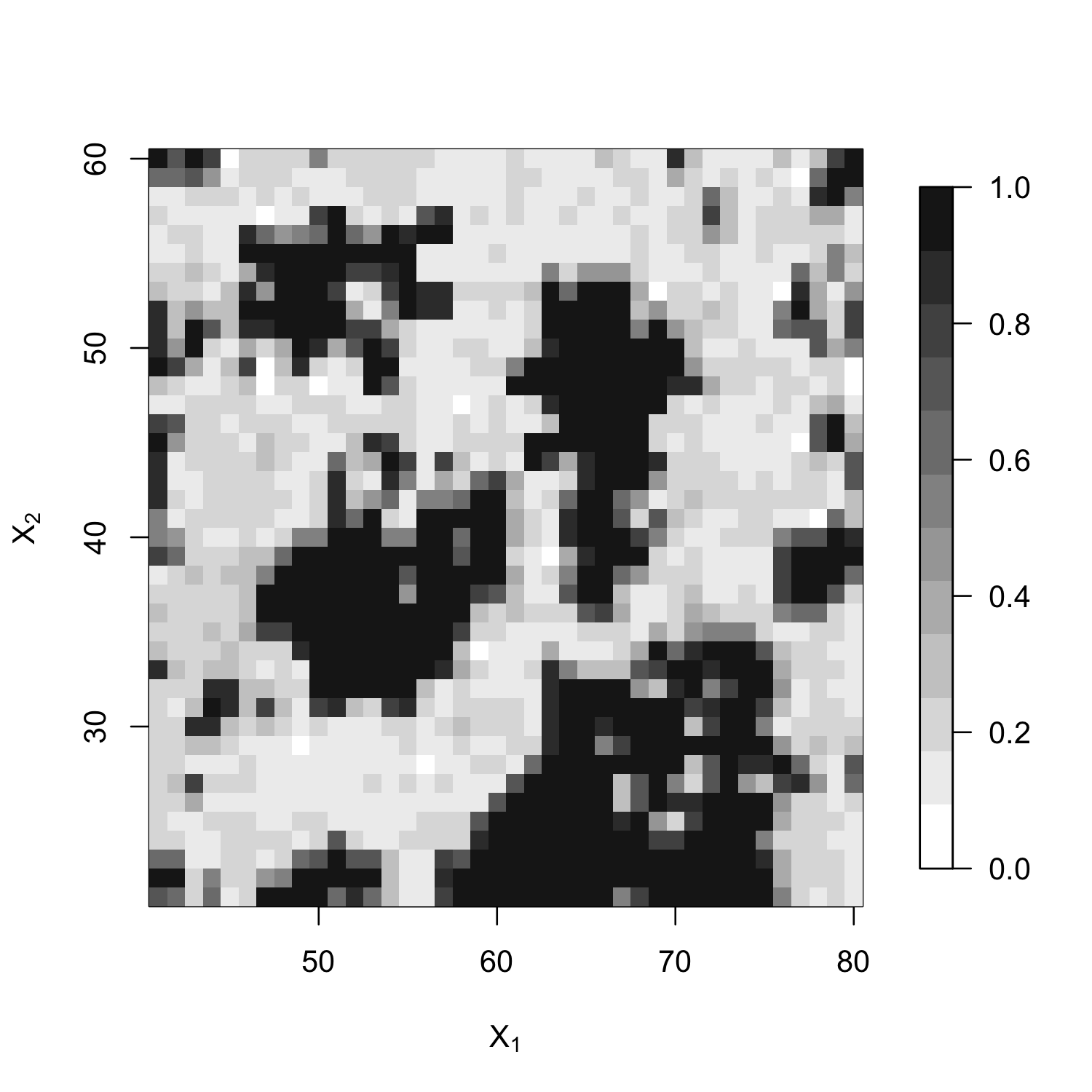}}
    \caption{Spatial model example: (a) original ice floe image with highlighted
    region. (b) close-up of focused region.}
    \label{IsingModel}
\end{figure}

We finish our examples by identifying ice floes from polar satellite images as
described in \citet{Banfield1992a}.  Here the image under consideration is a
$200$ by $200$ gray-scale satellite image, with focus on a particular $40$ by
$40$ region ($y$, Figure \ref{IsingModel}); the goal is to identify the
presence and position of polar ice floes ($x$).  Towards this goal,
\citet{Higdon1998b} employs a Bayesian model.  Basing the likelihood on
similarity to the image and employing an Ising model prior, the resulting
posterior distribution is
\begin{align*}
	\log( \pi(x | y) ) \propto \alpha \sum_i I [ y_i = x_i ] + \beta \sum_{i \sim j} I [ x_i = x_j ].
\end{align*}
The first term, the likelihood, encourages states $x$ which are similar to the
original image $y$.  The second term, the prior, favors states $x$ for which
neighbouring pixels are equal.  Here neighbourhood ($\sim$) is defined as the
$8$ vertical, horizontal, and diagonal adjacencies of each (interior) pixel.

Because the prior strongly prefers large blocks of identical pixels, an MCMC
method which proposes to flip one pixel at a time will fail to explore the
posterior, and hence \citet{Higdon1998b} suggests a partial decoupling
technique specific to these types of models.  However, to demonstrate PAWL's
power and universality, we demonstrate its ability to make simple one-at-a-time
Metropolis-Hastings feasible in these models without more advanced decoupling
methods.

\begin{figure}
    \centering
    \includegraphics[width=0.95\textwidth]{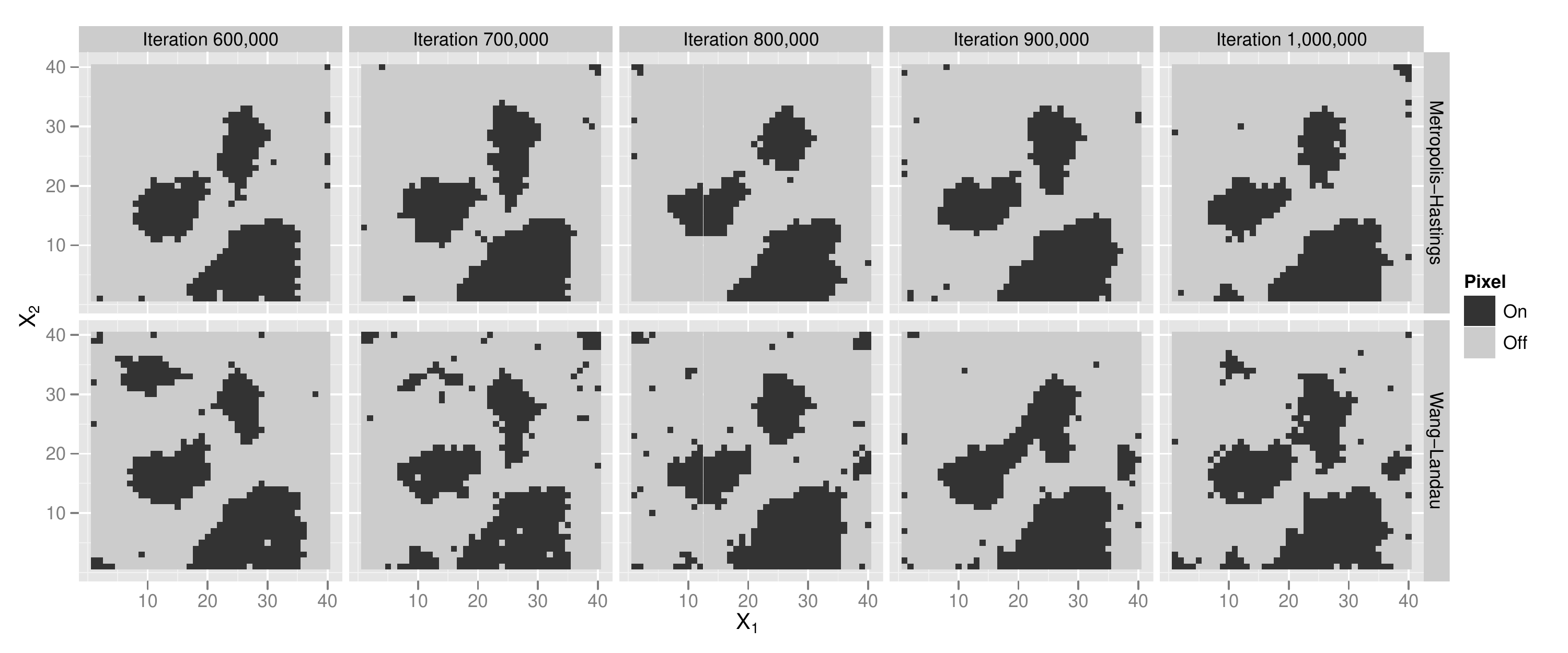}
    \caption{Spatial model example: states explored over 400,000 iterations for
    Metropolis-Hastings (top) and proposed algorithm (bottom).}
    \label{SpatialExplore}
\end{figure}

\begin{figure}
    \centering
    \includegraphics[width=0.95\textwidth]{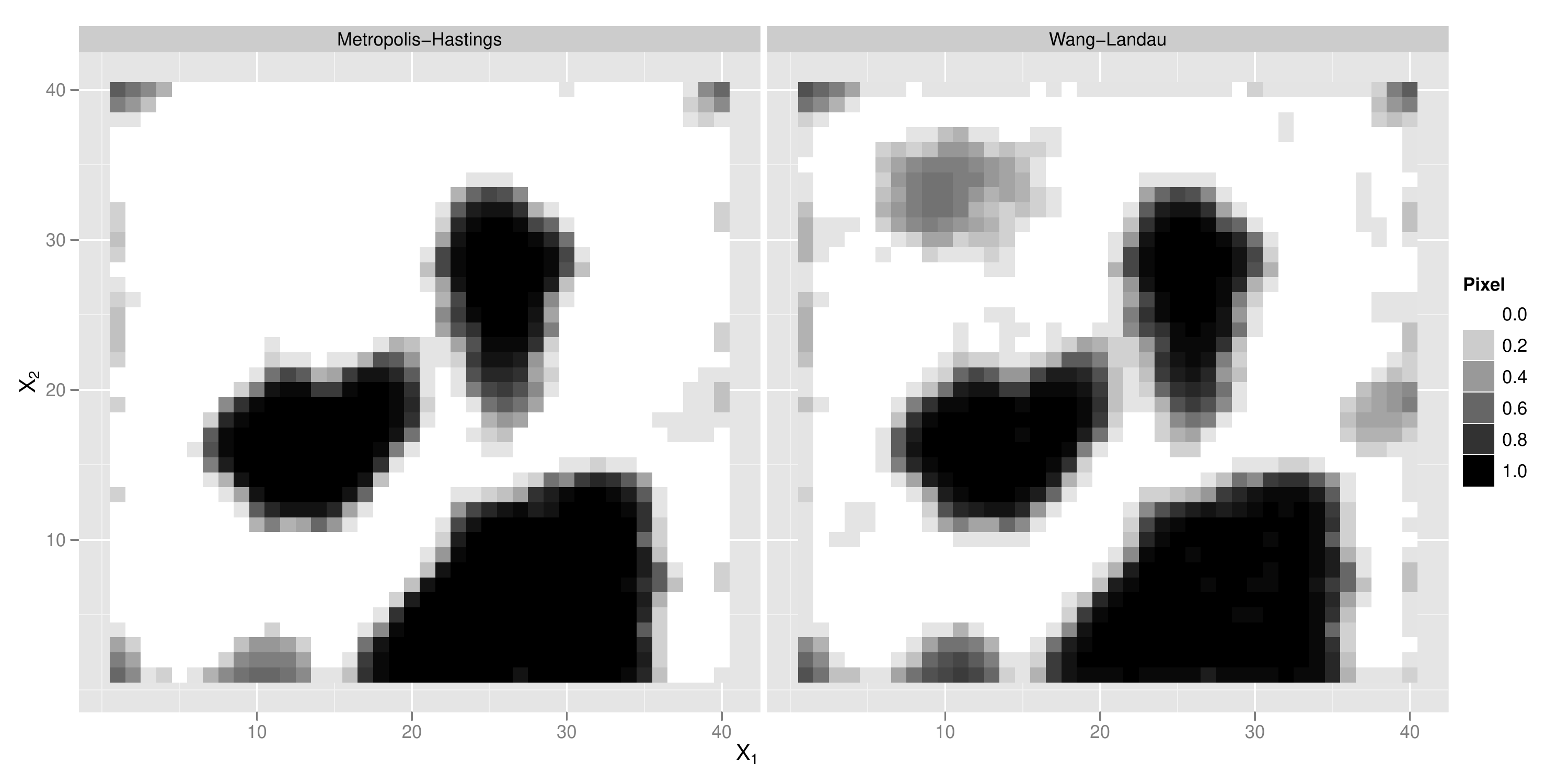}
    \caption{Spatial model example: average state explored with Metropolis-Hastings
    (left) and PAWL after importance sampling (right).}
    \label{SpatialMeanExplore}
\end{figure}

First running a preliminary Metropolis-Hastings algorithm of length $20,000$,
we use the range of explored energy values divided evenly across $10$ bins.
The algorithm subsequently splits bins 6 times (with splitting stopped once the
algorithm reaches the extremes of the reaction coordinate values) resulting in
$17$ bins at the algorithm's conclusion.  For both algorithms, we run $10$
chains for $1,000,000$ iterations with model parameters $\alpha=1$,
$\beta=0.7$.  Due to the flip-one-pixel approach, we suppress adaptive
proposals for this example.  In contrast to the mixture modeling example, in
this example the target density is fairly straightforward to calculate, so it
is a good worst-case comparison to demonstrate the additional time taken by the
proposed algorithm.  For this example, the MH algorithm took $388 \pm 21$
seconds across 10 runs, whereas PAWL required $478 \pm 24$ seconds.  Thus in
this case the Wang-Landau adds a 23\% price to each iteration on average.
However, as we will show, the exploration is significantly better, justifying
the slight additional cost.  Figure \ref{SpatialExplore} shows a subset of the
last $400,000$ posterior realizations from one chain of each algorithm.  We see
that the proposed Wang-Landau algorithm encourages much more exploration of
alternate modes.  The corresponding average state explored over all $10$ chains
(after $400,000$ burn-in) is shown in Figure \ref{SpatialMeanExplore}.  From
this we see that Wang-Landau induces exploration of the mode in the top-left of
the region in question, as well as a bridge between the central ice floes.  In
conclusion, while flip-one-pixel Metropolis-Hastings is incapable of exploring
the modes in the posterior caused by the presence/absence of large ice floes,
the proposed algorithm encourages exploration of these modes, even in the
presence of high between-pixel correlation.  While \citet{Higdon1998b} develops
a custom-tailored MCMC solution to overcome the inability of
Metropolis-Hastings to adequately explore the posterior density in Ising
models, we employ PAWL -- a general-purpose automatic density exploration
algorithm -- to achieve similar results.

\section{Discussion and Conclusion}

The proposed algorithm, PAWL, has at its core the Wang-Landau algorithm which,
despite wide-spread use in the physics community, has only recently been
introduced into the statistics literature.  A well-known obstacle in
implementing the Wang-Landau algorithm is selecting the bins through which to
discretize the state space; in response, we have developed a novel adaptive
binning strategy.  Additionally, we employ an adaptive proposal mechanism to
further reduce the amount of user-defined parameters.  Finally, to improve the
convergence speed of the algorithm and to exploit modern computational power,
we have developed a parallel interacting chain version of the algorithm which
proves efficient in stabilizing the algorithm.  Through a host of examples, we
have demonstrated the algorithm's ability to conduct density exploration over a
wide range of distributions continuous and discrete.  While a suite of
custom-purposed MCMC tools exist in the literature for each of these models,
the proposed algorithm handles each within the same unified framework.

As practitioners in fields ranging from image-processing to astronomy turn to
increasingly complex models to represent intricate real-world phenomena, the
computational tools to approximate these models must grow accordingly.  In this
paper, we have proposed a general-purpose algorithm for automatic density
exploration.  Due to its fully adaptive nature, we foresee its application as a
black-box exploratory MCMC method aimed at practitioners of Bayesian methods.
While statisticians are well-accustomed to performing exploratory analysis in
the modeling stage of an analysis, the notion of conducting preliminary
general-purpose exploratory analysis in the Monte Carlo stage (or more
generally, the model-fitting stage) of an analysis is an area which we feels
deserves much further attention.  As models grow in complexity, and endless
model-specific Monte Carlo methods are proposed, it is valuable for the
practitioner to have a universally applicable tool to throw at their problem
before embarking on custom-tuned, hand-built Monte Carlo methods.  Towards this
aim, the authors have published an $\texttt{R}$ package ("PAWL") to minimize
user effort in applying the proposed algorithm to their specific problem.

\section*{Acknowledgements}

Luke Bornn is supported by grants from NSERC and The Michael Smith Foundation
for Health Research.  Pierre E. Jacob is supported by a PhD fellowship from the
AXA Research Fund.  The authors also acknowledge the use of \texttt{R} and
\texttt{ggplot2} in their analyses \citep{R-Development-Core-Team2010a,
Wickham2009a}.  Finally, the authors are thankful for helpful conversations
with Yves Atchad\'{e}, Fran\c{c}ois Caron, Nicolas Chopin, Faming Liang, Eric
Moulines, and Christian Robert.

\setstretch{1.4}
\bibliographystyle{apalike}
\bibliography{lukebornnmodified}

\setstretch{1.48}

\appendix

\section{Details of Proposed Algorithm}

In Algorithm \ref{algo:final} we detail PAWL, fusing together a Wang-Landau base with adaptive binning,
interacting parallel chains, and an adaptive proposal mechanism.  In comparison
to the generalized Wang-Landau algorithm (Algorithm \ref{algo:generalizedWL}),
when a flat histogram is reached the distribution of particles within bins is
tested to determine whether a given bin should be split.  In addition, a suite
of $N$ interacting chains is employed, and hence the former chain $\bm{X}_t$ is
now made of $N$ chains: $\bm{X}_t = (X_t^{(1)}, \ldots, X_t^{(N)})$, each
defined on the state space $\mathcal{X}$. All the $N$ chains are used to update
the bias $\theta_t$, as described in Section \ref{sec:parallel}.

The chains are moved using an adaptive mechanism determined by the
Metropolis-Hastings acceptance rate as explained in Section
\ref{sec:proposals}.  While we present Algorithm \ref{algo:final} with adaptive
proposal variance, it may also be implemented with an adaptive mixture proposal
as described in Section \ref{sec:proposals}.  Note that when a bin is split, it
is possible to set the desired frequency of the new bins to some reduced value,
say each obtaining half the desired frequency of the original -- in fact in the
numerical experiments we do exactly that.  However, for notational and
pedagogical simplicity, we present here the algorithm where the desired
frequency of each bin is equal to $1/d_t$ at iteration $t$.

\setstretch{1.15}
\begin{algorithm}
    \caption{Proposed Density Exploration Algorithm}
    \begin{algorithmic}[1]
        \STATE Run a preliminary exploration of the target e.g. using adaptive MCMC, and
        determine an energy range.
        \STATE Partition the state space into $d_0$ regions $\{ \mathcal{X}_{1,0}, \dots,
        \mathcal{X}_{d_0,0} \}$ along a reaction coordinate $\xi(x)$,\\ the default choice
        being $\xi(x) = -\log \pi(x)$.
        \STATE $\forall i\in \{ 1,\dots, d_0 \}$ set $\theta(i)\leftarrow1,  \nu(i) \leftarrow 0$.
        \STATE Choose an initial proposal standard deviation $\sigma_0$
        \STATE Choose the frequency $\tau$ with which to check for a flat histogram.
        \STATE Choose a decreasing sequence $\{ \rho_t \}$, typically $\rho_t = 1 / t$, to update the proposal standard deviation.
        \STATE Choose a decreasing sequence $\{ \gamma_k \}$, typically $\gamma_k = 1/k$, to update the bias.
        \STATE Sample $\bm{X}_0 \sim \pi_0$, an initial distribution.
        \FOR {$t=1$ to $T$}
        \STATE Sample $\bm{X}_t$ from $P_{\theta_{t-1}}(\bm{X}_{t-1},\cdot)$, a
        transition kernel with invariant distribution
        $\prod_{n=1}^{N}\tilde{\pi}_{\theta_{t-1}}(x)$, parametrized by the
        proposal standard deviation $\sigma_{t-1}$.
        \STATE Update the proposal standard deviation: $\sigma_{t} \leftarrow
        \sigma_{t-1}+ \rho_t \left( 2 \mathcal{I}(A > .234) - 1 \right)$,\\
        where $A$ is the last acceptance rate.
        \STATE Set $d_{t} \leftarrow d_{t-1}$.
        \STATE Update the proportions: $\forall i\in \{1,\dots,d_{t}\} \quad \nu(i)
        \leftarrow \frac{1}{t}\left[ (t-1) \nu(i) + N^{-1} \sum_{j=1}^N
        \mathcal{I}_{\mathcal{X}_{i,t}}(X_t^{(j)})\right]$.
        \STATE Every $\tau$-th iteration, check the distribution of samples within each bin, extending the range if necessary.
        For example, if if $\xi(x) = -\log \pi(x)$ and a new minimum value of $\xi(x)$ was found, extend the first bin in order to include
        this value.
        \FOR {$i \in \{1, \dots, d_t\}$}
        \IF {bin $i$ should be split}
        \STATE Create two sub-bins covering bin $i$, assign to each a weight equal to $\theta_{t}(i)/2$.
        \STATE Set $d_{t} \leftarrow d_{t} + 1$, extend $\nu$.
        \ENDIF
        \ENDFOR
        \IF {``flat histogram'': $\max_{i \in \{1,\dots,d_{t}\}} \vert \nu(i) - {d_{t}}^{-1}\vert < c/d$}
        \STATE Set $k \leftarrow k + 1$.
        \STATE Reset $\forall i\in \{1,\dots,d_{t}\} \quad \nu(i) \leftarrow 0$.
        \ENDIF
        \STATE Update the bias: $\log \theta_{t}(i) \leftarrow \log \theta_{t-1}(i)  + \gamma_k
        (N^{-1} \sum_{j=1}^N \mathcal{I}_{\mathcal{X}_{i,t}}(X_t^{(j)}) - d_t^{-1})$.
        \STATE Normalize the bias: $\theta_t(i) \leftarrow \theta_t(i) / \sum_{i = 1}^{d_t} \theta_t(i)$.
        \ENDFOR	
    \end{algorithmic}
    \label{algo:final}
\end{algorithm}
\setstretch{1.48}

\section{Algorithm Convergence}

The convergence of the Wang-Landau algorithm using a deterministic stepsize,
also called the Stochastic Approximation Monte Carlo (SAMC) algorithm, and
stochastic approximation algorithms in general, has been well-studied in
\cite{Atchade2010a, Andrieu2006a, Andrieu2006b, Liang2011a}, see Chapter 7 of
\cite{Liang2010a} for a recent introduction. Since writing this manuscript, we
have also learned of recent convergence and ergodicity results for a parallel
implementation of the SAMC algorithm \citep{Liang2011a}. However, as noted in
\cite{Liang2011a} these results fail to explain why the parallel version is
more efficient than the single-chain algorithm in practice; instead it proves
the consistency of the algorithm when the number of iterations goes to
infinity, and the asymptotic normality of the bias $(\theta_t)_{t\geq 0}$, for
any fixed number of chains. We believe that precise statements on the impact of
the number of chains upon  the stabilization of the bias $(\theta_t)_{t\geq 0}$
would require the analysis of the Feynman--Kac semigroup associated with the
algorithm, similar to what is commonly used to study the impact of the number
of particles in Sequential Monte Carlo methods \citep{delmoral:book}.

Each of our proposed improvements adds a level of complexity to the proof of
the algorithm's consistency.  First and foremost, we are using the Flat
Histogram criterion, and thus the usual assumptions on the stepsize of the
stochastic approximation are not easily verified (e.g. assumptions of Theorem
2.3 in \cite{Andrieu2006b} and conditions (A4) in \cite{Liang2011a}). Indeed,
if no flat histogram criterion was met, then the stepsize $(\gamma_k)_{k\geq
0}$ would stay constant. We rely on a result in \cite{jacob:ryder:2011} that
proves that the criterion is met in finite time, for any precision threshold
$c$; therefore the results of \cite{Andrieu2006b} and thus the results of
\cite{Liang2011a} apply even when one uses the flat histogram criterion.

Finally, with our inclusion of an adaptive proposal as a Robbins-Monro style
update, the algorithm still remains in the class of stochastic approximation
algorithms. One could pragmatically stop the adaptation of the proposal
distribution after some iteration and fall back to the study of a homogeneous
Metropolis--Hastings algorithm. However, we believe that the algorithm could be
studied in the same framework as \cite{Andrieu2006a, Liang2011a}, where now the
stochastic approximation would both control the bias $(\theta_t)_{t \geq 0}$
and the standard deviation of the proposal $(\sigma_t)_{t \geq 0}$. 

\section{Trimodal Target Example}

\begin{figure}
    \centering
    \includegraphics[width=0.46\textwidth]{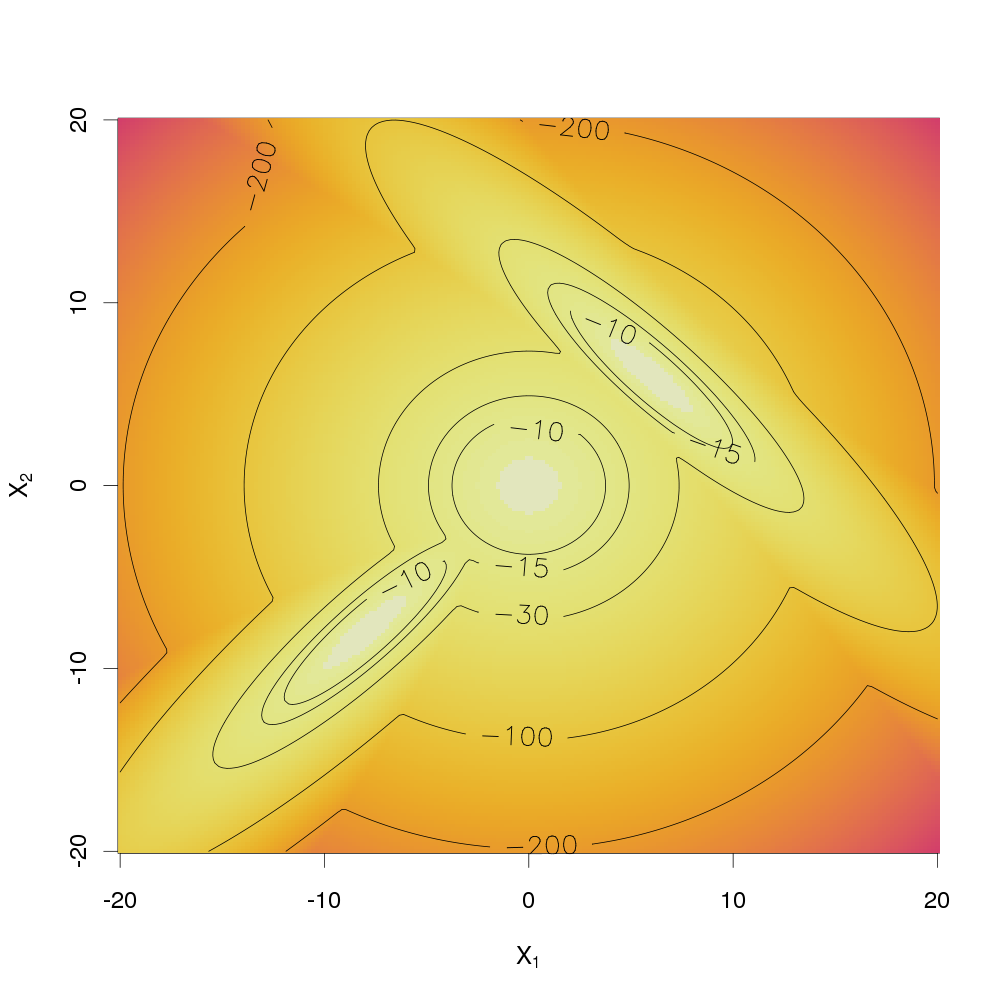}
    \caption{Trimodal example: log density function of the target
    distribution (\ref{trimodalDensity}). The modes are separated by areas where the
    log density is very low, making exploration difficult.}
    \label{trimodalLogDensity}
\end{figure}

\begin{figure}
    \centering
    \subfigure[Biased target density]{\label{subfig:biasedtarget}\includegraphics[width=0.46\textwidth]{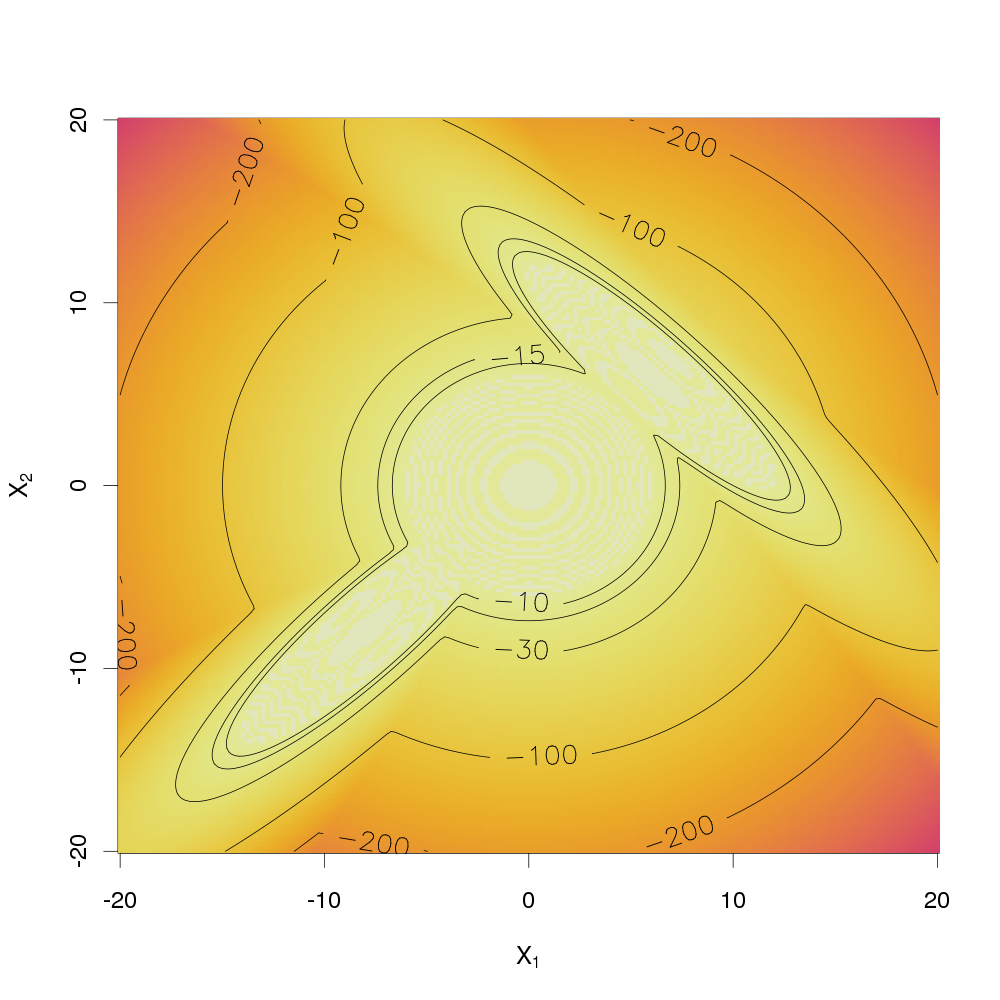}}
    \subfigure[Marginal of biased and unbiased target
    density]{\label{subfig:biasedmarginals}\includegraphics[width=0.46\textwidth]{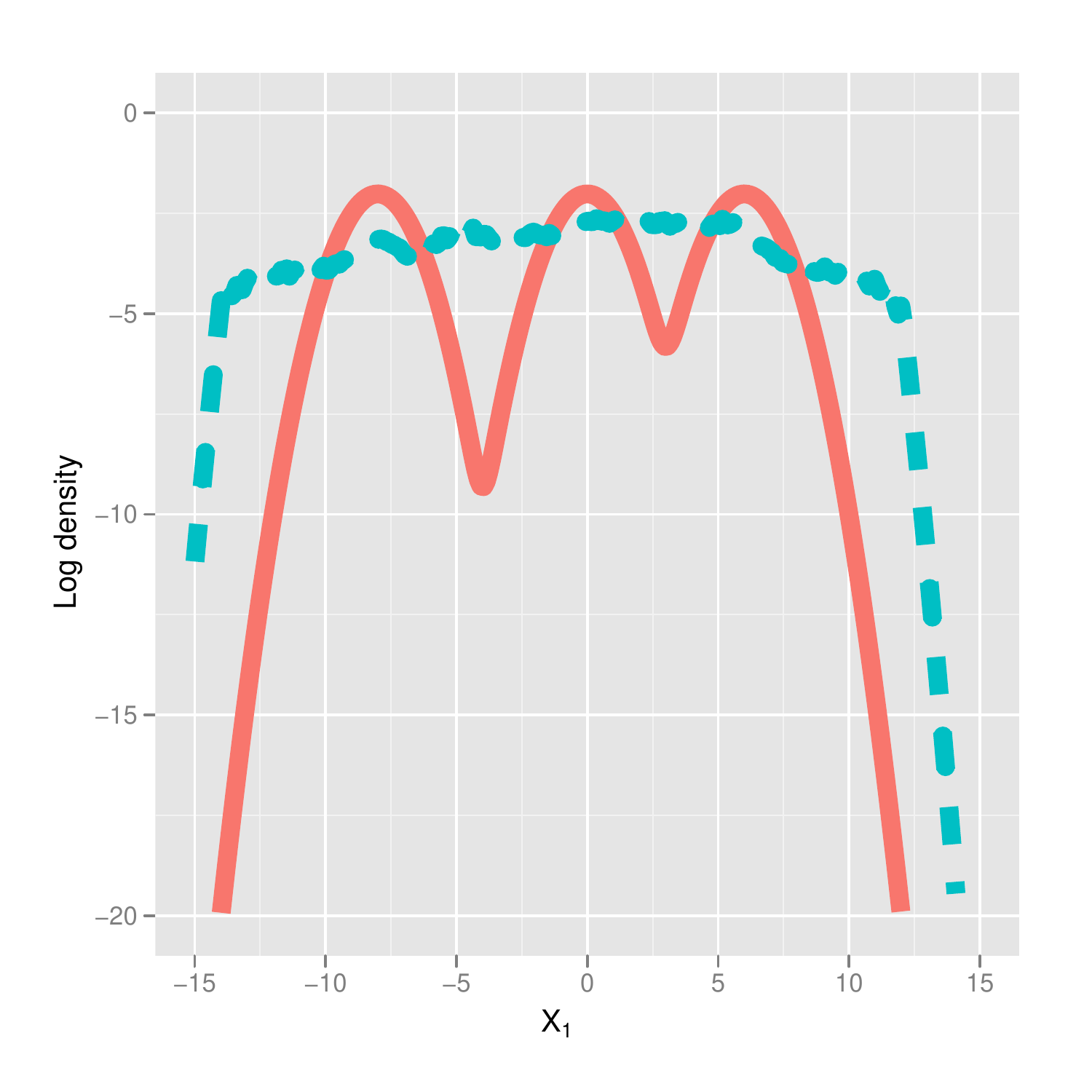}}
    \caption{Trimodal example: (left) log density of the biased target distribution
    (Equation (1)). The former modes are now all located in a fairly flat
    region, allowing for straightforward exploration.  (Right) marginal log density
    of one component of the (symmetric) trimodal target. The solid line shows the
    target probability density function and the dashed line shows an approximation
    of the marginal of the biased distribution of Equation (1).
    The biased marginal is flatter, hence easier to explore than the original target
    distribution.}
    \label{trimodalBiased}
\end{figure}

We introduce a toy target distribution to aid in demonstrating some
of the concepts discussed earlier, especially the bin splitting strategy.
Consider the 2-dimensional trimodal target
described in \citet{Liang2007a}:
\begin{align}
    X \sim \frac{1}{3} N \left[ \left( \begin{array}{c} 8 \\ 8 \end{array}
        \right) , \left( \begin{array}{cc} 1 & .9 \\ .9 & 1 \end{array} \right)
            \right]
        + \frac{1}{3} N \left[ \left( \begin{array}{c} 6 \\ 6 \end{array}
            \right) , \left( \begin{array}{cc} 1 & -.9 \\ -.9 & 1 \end{array}
                \right) \right]
            + \frac{1}{3} N \left[ \left( \begin{array}{c} 0 \\ 0 \end{array}
                \right) , \left( \begin{array}{cc} 1 & 0 \\ 0 & 1 \end{array}
                    \right) \right],
                \label{trimodalDensity}
\end{align}
a mixture of three bivariate Gaussian distributions.  The corresponding log
density is shown in Figure \ref{trimodalLogDensity}.  This density, while
low-dimensional and with only three modes, is known to be difficult to sample
from with Metropolis-Hastings (e.g. \citealt{Gilks1998a}).  Firstly, with
different correlation structures in each mode, an adaptive algorithm might
conform to one mode, missing (or poorly sampling from) the other two.
Secondly, there is a low-density region separating each mode; as such,
low-variance proposals might be incapable of jumping between modes.  Figure
\ref{trimodalBiased} displays the biased target (Equation (1), using the log
density as reaction coordinate) and one of its marginals, emphasizing the
effect of biasing in improving the ability for the algorithm to explore the
density.  Here the plot is created using computationally expensive fine-scale
numerical integration.

\begin{figure}
    \centering
    \subfigure[Scatter plot of the chains generated by the proposed
    algorithm]{\label{subfig:trimodal:scatterplotPAWL}\includegraphics[width=0.4\textwidth]{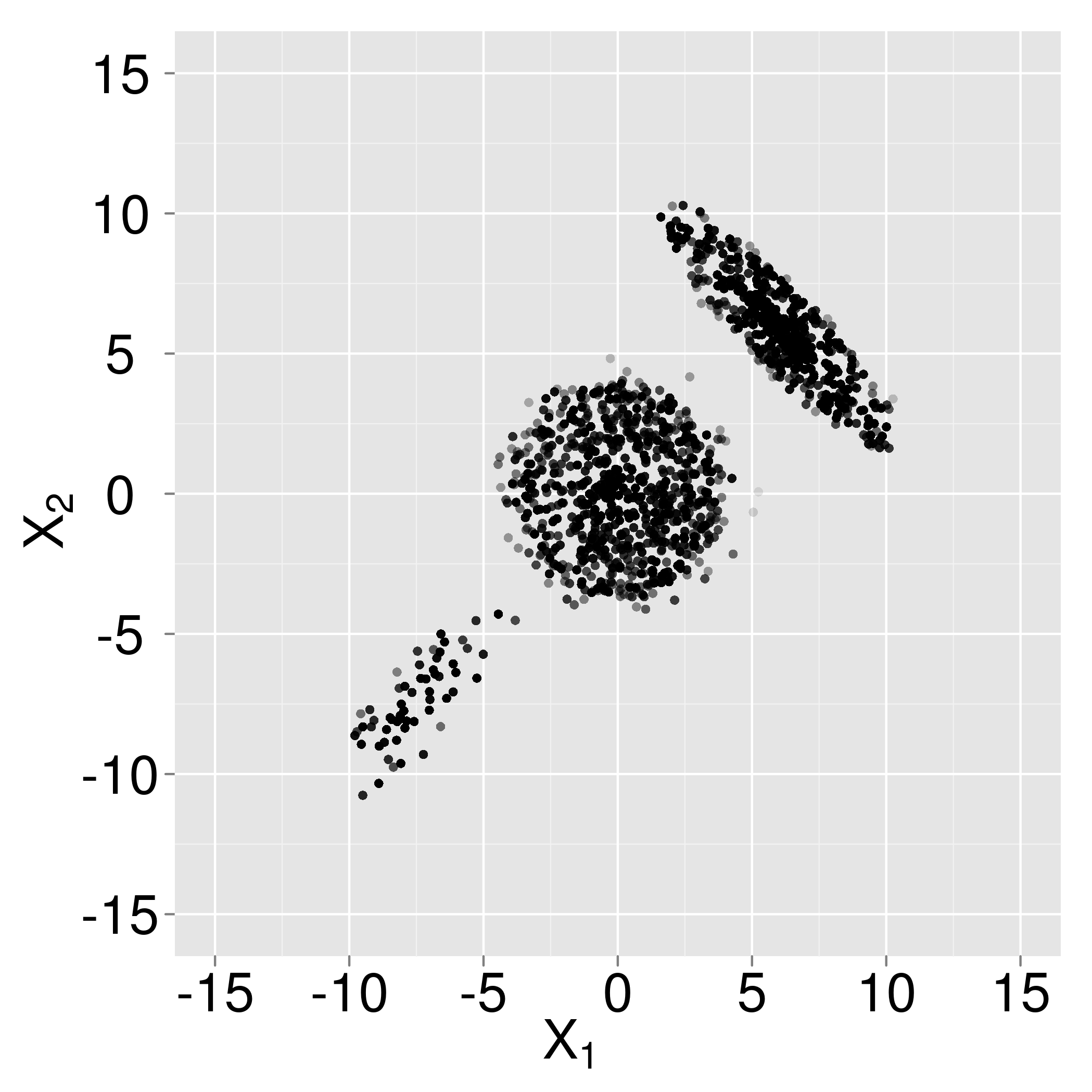}}
    \subfigure[Scatter plot of the chains generated by an adaptive
    Metropolis--Hastings algorithm]{\label{subfig:trimodal:scatterplotAMH}
    \includegraphics[width=0.4\textwidth]{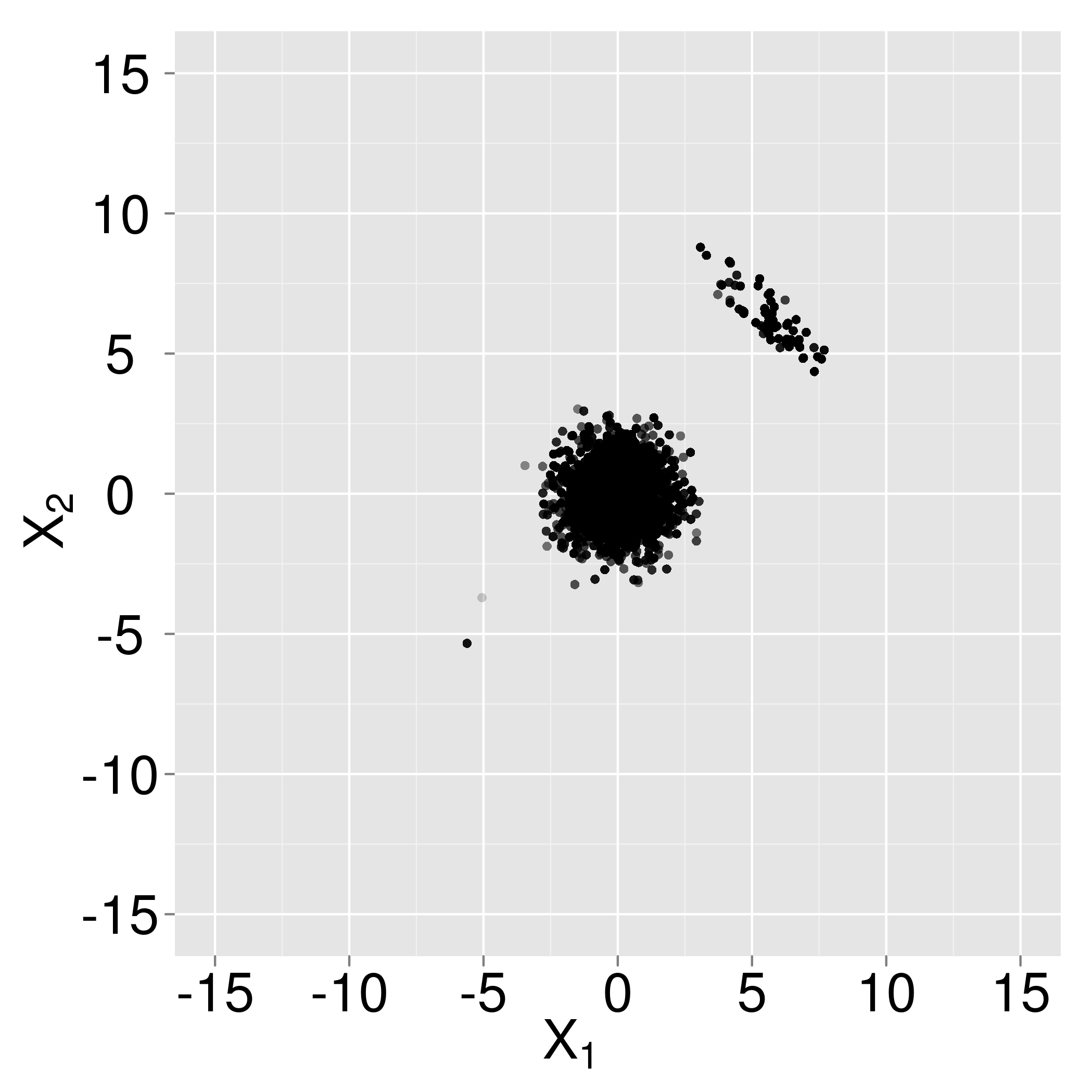}}
    \caption{\label{fig:trimodal:results} Trimodal example: results of the proposed
    algorithm: (a) Scatter plot of all samples, before normalization/importance
    sampling, (b) Scatter plot of samples generated by an adaptive
    Metropolis--Hastings algorithm using the same number of chains and iterations.}
\end{figure}

Initial exploration is performed by an adaptive MCMC algorithm, run with $2$
parallel chains and $500$ iterations. The proposal of this first run targets a
specific acceptance rate of $23.4\%$, as described in Section 3.4. The explored
energy range is expanded and divided into $d_0 = 3$ initial bins.  In all
examples, we use $c=0.5$.  The proposed algorithm is run with $2$ parallel
chains for $2500$ iterations. Figure \ref{subfig:trimodal:scatterplotPAWL}
shows the regions recovered by the chains; the chains have moved easily between
the modes, even if the distribution of the starting point was not well spread
over the target density support. In this case, to reflect the possible lack of
information about the target support, we draw the starting points of all the
chains from a $\mathcal{N}(0, 0.1\times I_2)$ distribution, hence exclusively
in one of the three modes.  In this setting the free energy SMC method
described in \citet{Chopin2010b} fails to recover the target distribution
accurately; specifically, the central mode is over-sampled due to many
particles not reaching the outer modes.  However, if the initial distribution
$\pi_0$ is well spread over the target support, the SMC algorithm recovers the
modes.  Figure \ref{subfig:trimodal:scatterplotAMH} shows the points generated
by $3000$ iterations of the adaptive Metropolis--Hastings algorithm (already
used in the initial exploration), also using $2$ chains. We see that the
exploration was less successful, with the bottom left mode hardly visited at
all, although the same number of point-wise density evaluations were performed
as for the proposed algorithm.

\begin{figure}
    \centering
    \subfigure[Trace plot of $\log \theta$, the log penalties, with vertical lines
    indicating the bin
    splits.]{\label{subfig:trimodal:logtheta}\includegraphics[width=\textwidth]{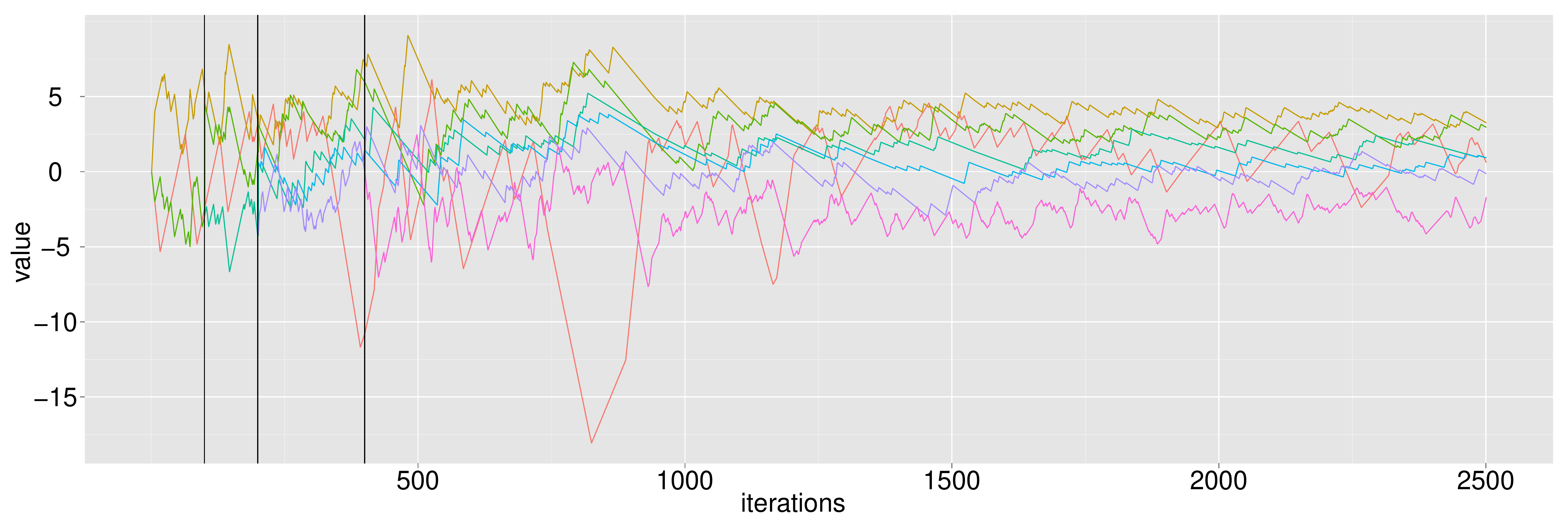}}
    \subfigure[Histogram of the energy values computed during the algorithm.
    Vertical full lines show the initial bins and dashed lines show the cuts that
    have been added dynamically.]{\label{subfig:trimodal:binhistogram}
    \includegraphics[width=\textwidth]{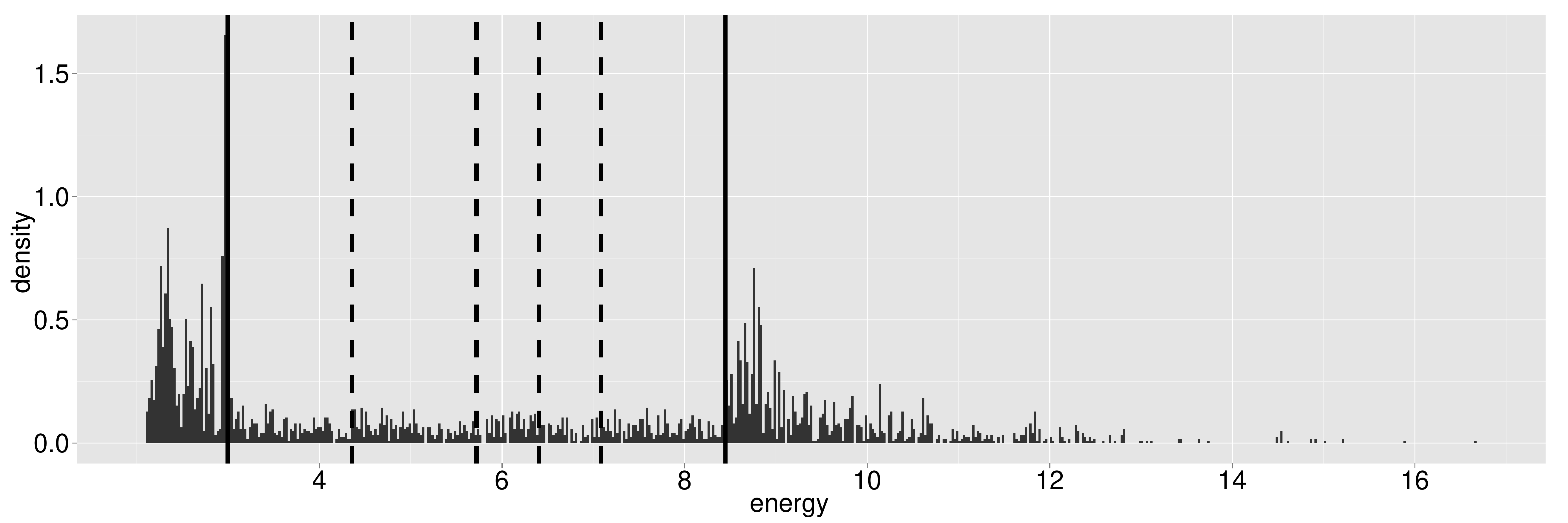}}
    \caption{Trimodal example: histograms of the log density values of all the
    chain points just before the iterations at which the splitting mechanism is
    triggered. The number of bins increases automatically along the iterations.
    \label{fig:trimodal:binsplit}}
\end{figure}

Along the iterations, the bins have been split three times. Here the chosen
strategy was to split a bin if less than $25\%$ of its points were situated in
half of the bin.  Figure \ref{fig:trimodal:binsplit} illustrates the effects of
the binning strategy. Figure \ref{subfig:trimodal:logtheta} shows the trace
plot of the estimators $\theta$, and the iterations at which bins are split are
shown with vertical lines. After each split, the dimension of $\theta$
increases but the figure shows that the new estimators quickly stabilize. After
the last split around iteration $450$, the number of bins stays constant.
Figure \ref{subfig:trimodal:binhistogram} shows the histogram of the log
density evaluations of the chain points, with vertical full lines showing the
initial bins, and vertical dashed lines showing the bins that have been added
during the run. We see that the bin splits induce more uniformity within bins,
and hence across the entire reaction coordinate range, aiding in movement and
exploration.

\end{document}